\documentclass[%
 reprint,
 amsmath,amssymb,
pra,
]{revtex4-2}

\usepackage[dvipdfmx]{graphicx}
\usepackage{dcolumn}
\usepackage{bm}
\usepackage{braket}
\usepackage{color}

\begin{document}
\preprint{APS/123-QED}

\title{Spectroscopic observation of the crossover from a classical Duffing oscillator to a Kerr parametric oscillator}

\author{T.~Yamaji$^{1, 2, 3}$}
	\thanks{These authors contributed equally to this work.}
\author{S.~Kagami$^{1, 2}$}
	\thanks{These authors contributed equally to this work.}
\author{A.~Yamaguchi$^{1, 2}$}
\author{T.~Satoh$^{1, 2, 3}$}
\author{K.~Koshino$^4$}
\author{H.~Goto$^5$}
\author{Z.~R.~Lin$^{3}$}
	\thanks{The current affiliation is Shanghai Institute of Microsystem and Information Technology (SIMIT), 865 Changning Road, Shanghai 200050, China.}
\author{Y.~Nakamura$^{3, 6}$}
\author{T.~Yamamoto$^{1, 2, 3}$}

\affiliation{$^1$System Platform Research Laboratories, NEC Corporation, Kawasaki, Kanagawa 211-0011, Japan}
\affiliation{$^2$NEC-AIST Quantum Technology Cooperative Research Laboratory, National Institute of Advanced Industrial Science and Technology (AIST), Tsukuba, Ibaraki 305-8568, Japan}
\affiliation{$^3$RIKEN Center for Quantum Computing (RQC), Wako, Saitama 351-0198, Japan}
\affiliation{$^4$College of Liberal Arts and Sciences, Tokyo Medical and Dental University, Ichikawa, Chiba 272-0827, Japan}
\affiliation{$^5$Corporate Research and Development Center, Toshiba  Corporation, Kawasaki, Kanagawa 212-8582, Japan}
\affiliation{$^6$Research Center for Advanced Science and Technology (RCAST), The University of Tokyo, Meguro, Tokyo 153-8904, Japan}
\date{\today}

\begin{abstract}
We study microwave response of a Josephson parametric oscillator consisting of a superconducting transmission-line resonator with an embedded dc-SQUID.
The dc-SQUID allows to control the magnitude of a Kerr nonlinearity over the ranges where it is smaller or larger than the photon loss rate. 
Spectroscopy measurements reveal the change of the microwave response from a classical Duffing oscillator to a Kerr parametric oscillator in a single device. 
In the single-photon Kerr regime, we observe parametric oscillations with a well-defined phase of either $0$ or $\pi$, 
whose probability can be controlled by an externally injected signal.
\end{abstract}

\maketitle

\section{INTRODUCTION}
A parametric oscillator is a driven oscillator, 
in which the resonance frequency is modulated typically at twice of it with a modulation amplitude larger than a threshold~\cite{Nayfeh_book, Strogatz_book}. 
This modulation called a parametric pump induces a self-oscillating state, which can be either of two states with an equal amplitude and a phase of $0$ or $\pi$. 
The amplitude of the self-oscillating state is usually saturated by dissipation such as one- and two-photon losses or nonlinearity such as the Kerr effect, where the Kerr effect shifts the resonance frequency as the amplitude increases. 
Parametric oscillators are implemented and utilized in a variety of physical systems~\cite{Giordmaine1965, Turner1998, Wilson2010}, especially in the classical regime, 
where the nonlinearity is much smaller than the photon loss rate out of the oscillator mode. 
For example, parametric oscillators formed by a nonlinear LC circuit were used as a building block of digital computers known as a parametron~\cite{Goto1959}, 
and optical parametric oscillators formed by a nonlinear crystal are widely used as a tunable light source~\cite{Boyd_book}. 
Furthermore, parametric oscillators operated below the pump threshold serve as a parametric amplifier, 
which is now an indispensable tool in the field of superconducting quantum information processing~\cite{Beltran2008}. There are also superconducting parametric oscillators with different operational modes such as nondegenerate oscillations~\cite{Wustmann2017, Bengtsson2018}, period-tripling subharmonic parametric oscillations~\cite{Svensson2017, Svensson2018}.

Recently, there has been growing interest in a parametric oscillator in the single-photon Kerr regime, a Kerr parametric oscillator (KPO)~\cite{Goto2016_scirep}, where the magnitude of the Kerr nonlinearity is larger than the photon loss rate and therefore the Kerr effect is visible at the single-photon level~\cite{Kirchmair2013}.
The KPOs have a wide range of potential applications such as deterministic generation of Schr\"{o}dinger cat state by adiabatic pumping in a single KPO~\cite{Goto2016_scirep, Puri2017, Goto2019} and quantum computation in a network of KPOs~\cite{Goto2016_scirep, Goto2016_pra, Puri2017, Nigg2017, Puri2017b, Zhao2018, Goto2019_jpsj, Goto2020, Kanao2020}. 
This regime has also been studied experimentally using trapped ions~\cite{Ding2017} and superconducting circuits~\cite{Wang2019, Grimm2020}. In the latter system, a Kerr-type nonlinearity is realized with Josephson junctions. 
Superconducting resonators with a Kerr nonlinearity larger than the photon loss rate (but not as large as those of transmon qubits) have also been studied 
from the view points of quantum simulation~\cite{Leib2012}, switching between dynamical states~\cite{Muppalla2018, Andersen2020} and an error-protected qubit encoded in an oscillator~\cite{Grimm2020}. 
As pointed out in Ref.~\citenum{Andersen2020}, this regime is relatively unexplored, and understanding the crossover between the classical and quantum regimes is important both in the theoretical and application points of view. 

In the present paper, we study microwave response of a superconducting transmission-line resonator with a Kerr nonlinearity realized by an embedded dc-SQUID, which works as a parametric oscillator called as a Josephson parametric oscillator (JPO). 
We design the device such that the magnitude of the Kerr nonlinearity can be smaller or larger than the photon loss rate depending on the flux bias for the dc-SQUID, 
which allows us to observe in a single device a significant difference in the microwave response between the classical Duffing-oscillator regime and the KPO regime. 
In the single-photon Kerr regime, we also observe parametric oscillations with a well-defined phase of either $0$ or $\pi$, 
which can potentially be utilized for applications in quantum information processing. 

\begin{figure}
\includegraphics[width=0.75\linewidth]{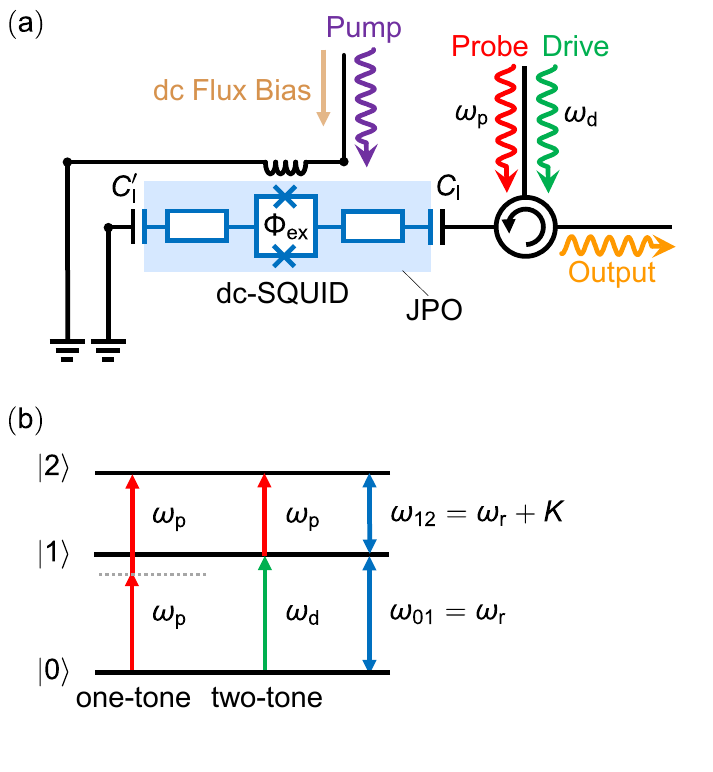}
\caption{(a) Schematic circuit diagram. The blue elements show the JPO, where two rectangles next to the dc-SQUID represent quarter-wavelength CPWs. 
(b) Energy-level diagram of the JPO, illustrating the schemes to determine the Kerr nonlinearity $K$ experimentally  in the single-photon Kerr regime.  Only the lowest three levels are shown. 
The red and green arrows show the probe~($\omega_{\rm p}$) and drive~($\omega_{\rm d}$) tones used in the spectroscopy measurements, respectively, where $\omega_{\rm d}$ is fixed at $\omega_{01}$ and $\omega_{\rm p}$ is swept. The left row of arrows depicts the two-photon absorption process at $\omega_\mathrm{p} = \omega_{02}/2$ in the one-tone spectroscopy, while center row of arrows shows the resonant transition at $\omega_{\rm p}=\omega_{12}$ in the two-tone spectroscopy.}
\label{fig1}
\end{figure}

\section{device}
Figure~\ref{fig1}(a) shows the circuit diagram of the JPO studied in this paper.
The JPO is composed of a half-wavelength coplanar waveguide (CPW) resonator interrupted by a dc-SQUID at the center of the resonator. 
The critical current $I_{\rm c}$ of each Josephson junction in the SQUID is estimated to be 1.0~$\mu$A 
from the room-temperature resistance of a test structure on the same chip.  
The CPW resonator is designed to be 4.614-mm long and is coupled to a feed line via an input capacitor $C_{\rm I}$ of 3.4~fF. The other end of the resonator is connected to the ground via a shunt capacitor $C_{\rm I}^\prime$ of 3.3~fF.
The dc-SQUID is inductively coupled to a pump line, through which a dc current and a pump microwave are applied 
to control the resonance frequency of the JPO and to induce a parametric oscillation, respectively (see Appendix A for more details on the device). 
The JPO is cooled below 10~mK in a dilution refrigerator. 

The resonance frequency $\omega_{\rm r}$ and the external and internal loss rates $\kappa_\mathrm{e}$ and $\kappa_\mathrm{i}$ of the JPO 
are evaluated by fitting the reflection coefficient of the JPO with the input--output theory \cite{Yamamoto_book}. At almost zero flux bias, $f \equiv \Phi_{\rm ex}/\Phi_{\rm 0}=0.00$,
where $\Phi_{\rm ex}$ is magnetic flux in the SQUID loop and $\Phi_{\rm 0}$ is the flux quantum, we obtain $\omega_\mathrm{r}/2\pi = 11.742$~GHz, $\kappa_\mathrm{e}/2\pi = 0.85$~MHz and $\kappa_\mathrm{i}/2\pi = 1.01$~MHz. 

The JPO behaves as a nonlinear oscillator described by the following Hamiltonian, 
\begin{equation}~\label{H_JPO}
\mathcal{H}/\hbar = \omega_{\rm r}(f) \, a^\dag a + \frac{K(f)}{2}a^\dag a^\dag a a,
\end{equation} 
where $a$ is the photon annihilation operator for the fundamental mode of the JPO, and
$\omega_{\rm r}(f)$ and $K(f)$ are respectively the resonance frequency and the negative Kerr nonlinearity periodically dependent on the flux bias. In the range $0<f<0.5$, $\omega_\mathrm{r}(f)$ decreases monotonically since the flux enhances the linear inductance of the SQUID, while $|K(f)|$ increases with $f$ due to the enhancement of the inductance participation ratio of the SQUID~\cite{Bourassa12}. The loss rates $\kappa_\mathrm{e}$ and $\kappa_\mathrm{i}$ are only weakly dependent on $f$.
We designed our JPO such that $|K|$ can be smaller or larger than the total loss rate $\kappa_{\rm tot}\equiv \kappa_{\rm e}+\kappa_{\rm i}$ depending on $f$. 

\section{spectroscopy measurements}
We evaluate the Kerr nonlinearity of the JPO by two types of measurements called one-tone and 
two-tone spectroscopies [Fig.~\ref{fig1}(b)].
In the one-tone spectroscopy, the reflection coefficient $\Gamma$ of the JPO 
is measured as a function of the frequency $\omega_{\rm p}$ (or detuning from the resonance frequency, $\Delta\omega_{\rm p}\equiv \omega_{\rm p}-\omega_{\rm r}$) and power $P_{\rm p}$ of the probe microwave by using a vector network analyzer. 
In the two-tone spectroscopy, another microwave tone called drive at a fixed frequency $\omega_{\rm d} = \omega_{\rm r}$ is added, 
and $\Gamma$ is measured as a function of the probe-tone frequency $\omega_{\rm p}$ (or $\Delta\omega_{\rm p}$) and the drive power $P_{\rm d}$. 
Hereinafter, the powers of various tones are specified at the relevant ports on the sample chip. 

In the Duffing-oscillator regime ($\left|K\right|<\kappa_{\rm tot}$), the energy-level discreteness is obscured by $\kappa$, and the Kerr nonlinearity can be deduced from probe-power-dependent continuous shifts of the resonance frequency~\cite{Yurke06}. 
In the single-photon Kerr regime ($\left|K\right|>\kappa_{\rm tot}$), on the other hand, the Kerr nonlinearity can be spectroscopically extracted as a discrete shift of the transition frequency as shown below, which is less sensitive to probe-power calibration compared to a method using ac Stark shift \cite{Frattini2018, Sivak2019}.

\begin{figure}
\includegraphics[width=\linewidth]{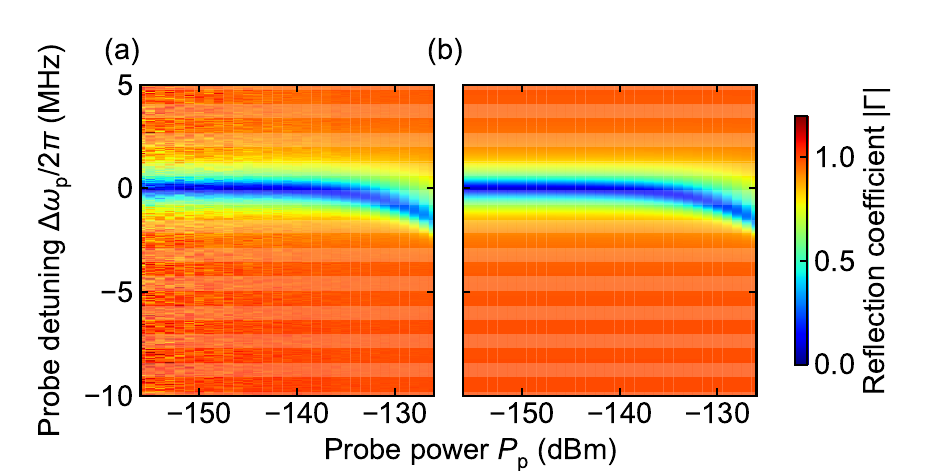}
\caption{One-tone spectra obtained in (a) the experiment at $f=0.00$ and (b) a simulation. The horizontal axes show the probe power $P_\mathrm{p}$. The vertical axes show $\Delta\omega_{\rm p}\equiv \omega_{\rm p}-\omega_{\rm r}$, the detuning of the probe tone from the resonance frequency of the JPO in the low probe-power limit ($P_\mathrm{p} = -156$~dBm). The color scale shows the amplitude of the reflection coefficient, which is normalized to unity in the off-resonant region. In the simulation, the Kerr nonlinearity $K(0.00)$ and the probe power $P_{\rm p}$ are treated as fitting parameters.}
\label{fig2}
\end{figure}

\begin{figure*}
\includegraphics[width=\linewidth]{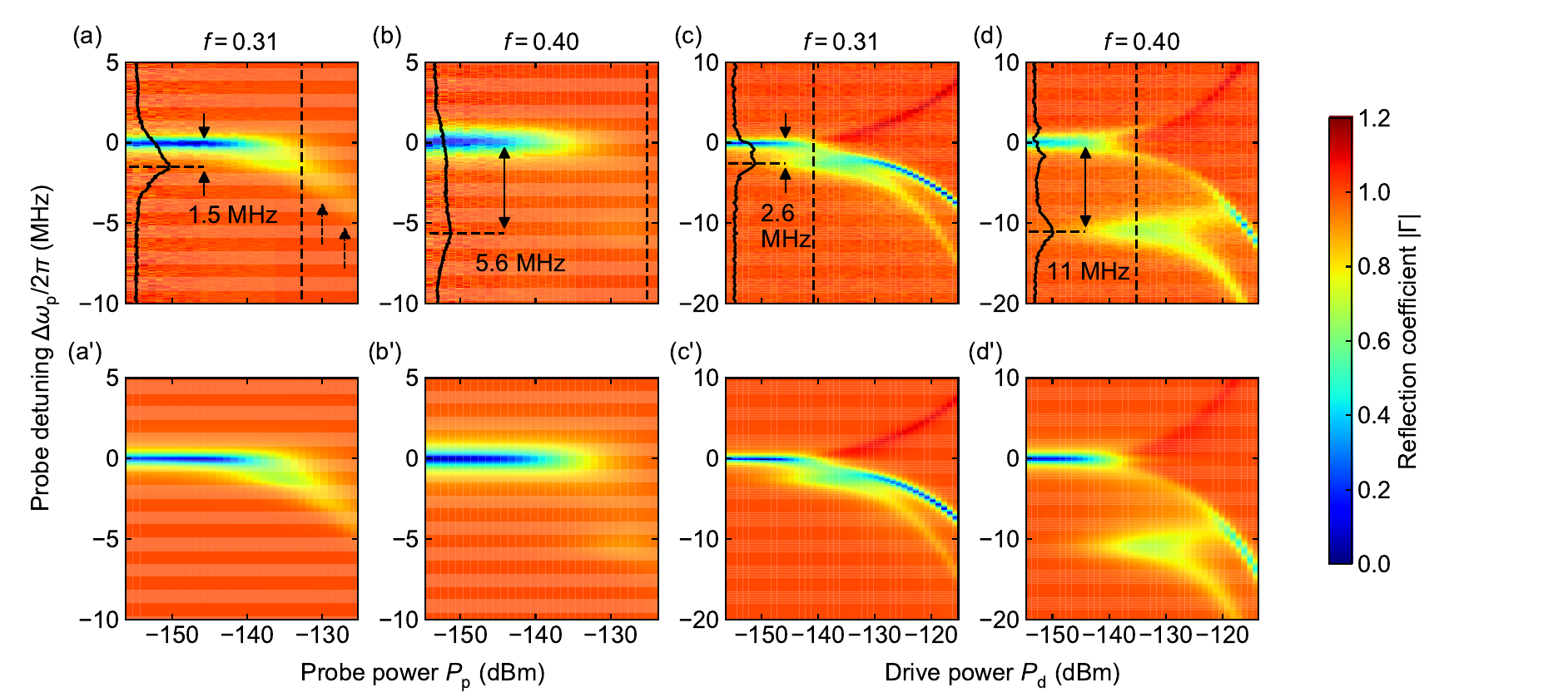}
\caption{
One-tone~(two-tone) spectroscopies at $f$=0.31 (a)~[(c)] and $f$=0.40 (b) [(d)]. 
Color scale shows the normalized amplitude of the reflection coefficient. 
The vertical axes represent the detuning of the probe tone 
from the resonance frequency in the low probe-power limit. 
The probe power in the two-tone measurements is set to $-145$~dBm.
The black trace in each figure shows the cross-section along the vertical dashed line, where the horizontal amplitude shows $|\Gamma|$ in a linearly descending unit. 
(a$^\prime$)--(d$^\prime$) are the numerical simulation corresponding to (a)--(d), respectively, where $K(0.31)/2\pi$~[$K(0.40)/2\pi$] are set to be the mean of $K$ determined from the spectroscopy measurements, $-2.8$~MHz~($-11$~MHz). 
}
\label{fig3}
\end{figure*}

\subsection{Spectroscopy under different flux biases}
Figure~\ref{fig2}(a) shows the one-tone spectrum measured at a flux bias close to zero ($f= 0.00$), where we see $\omega_{\rm r}$ decreases continuously and monotonically with $P_{\rm p}$.
This behavior is typical for a classical Duffing oscillator with a negative Kerr nonlinearity~\cite{Nayfeh_book, Strogatz_book}. 
The observed spectrum is well reproduced by the simulation shown in Fig.~\ref{fig2}(b), where we numerically determine the stationary amplitude of the JPO resonator with the Hamiltonian of Eq.\!~(\ref{H_JPO}) and calculate the reflection coefficient of the probe microwave based on the input--output theory (see Appendix~B for details). In the calculation, we used the parameters $\omega_{\rm r}$, $\kappa_{\rm e}$ and $\kappa_{\rm i}$, which are determined from the reflection coefficient measurement in the low probe-power limit, while $K$ and $P_{\rm p}$ are treated as fitting parameters. 
The fitted probe power is consistent with the measurement of the line attenuation within the precision of 2~dB. The Kerr nonlinearity $|K(0.00)|/2\pi$ is determined to be $0.45$~MHz, which is lower than the total loss rate $\kappa_\mathrm{tot}/2\pi = 1.86$~MHz at $f=0.00$ and is consistent with the observed behavior of a classical Duffing oscillator.

The magnitude of the Kerr nonlinearity can be enhanced by applying a flux bias to the SQUID. 
Figures~\ref{fig3}(a) and \ref{fig3}(c) show the spectra of the one-tone and two-tone spectroscopies at $f=0.31$, respectively, where $\omega_{\rm r}/2\pi$, $\kappa_\mathrm{e}/2\pi$ and $\kappa_\mathrm{i}/2\pi$ are measured to be 11.019~GHz, 0.57~MHz, and 0.34~MHz, respectively.
In the one-tone spectroscopy shown in Fig.~\ref{fig3}(a), we observe a dip at $\omega_{\rm r}$ in the weak power regime. The frequency and depth of the dip do not depend on the probe power because the mean photon number in the resonator is much smaller than unity, and the resonator is in the linear response regime.
The dip at $\omega_{\rm r}$ disappears due to absorption saturation as we increase $P_{\rm p}$ to $\sim-140$~dBm, where the population of the $\ket{1}$ state is comparable to that of the $\ket{0}$ state and the oscillator strength of the $\ket{0}$-to-$\ket{1}$ transition decreases to zero. 
At $P_{\rm p}\sim -140$~dBm, we observe the two-photon absorption for the transition between the Fock states $|0\rangle$ and $|2\rangle$ at $\omega_{02}/2=\omega_{\rm r}+K/2$ [Fig.~\ref{fig1}(a)] (The three and four-photon absorptions at $\omega_{\rm r}+K$ and $\omega_{\rm r}+3K/2$ are also visible in the higher power region [dashed arrows in Fig.~\ref{fig3}(a)]). 
We determine $|K(0.31)|/2\pi$ to be 3.0~MHz from the center frequency of the two-photon absorption dip in the power region where the two-photon absorption is stronger than the $\omega_{01}$ transition [vertical dashed line in Fig.~\ref{fig3}(a)]. The frequency of the two-photon absorption slightly decreases with $P_{\rm p}$ due to Rabi splitting caused by the mixing of $\ket{0}$, $\ket{1}$ and $\ket{2}$ states, which leads to the uncertainty in the estimation of the Kerr nonlinearity \cite{Koshino2013}. 

In the two-tone spectroscopy shown in Fig.~\ref{fig3}(c), the dip at $\omega_{\rm r}$ is also independent of drive power in the low-power regime ($P_{\rm d} \lesssim  -140$~dBm).
Contrary to the one-tone spectroscopy, the dip splits into two branches as we increase $P_{\rm d}$, where the strong drive mixes $\ket{0}$ and $\ket{1}$ states to form dressed states (Rabi splitting: see Appendix~D). The higher- and lower-frequency branches show amplification and attenuation of the probe, respectively. These are caused by the difference in the population of the dressed states \cite{Koshino2013}. 
In the region with $P_{\rm d}\gtrsim -140$~dBm, the drive tone populates the $\ket{1}$ state and the transition between $\ket{1}$ and $\ket{2}$ states is directly observed as an absorption at $\omega_{12}=\omega_{\rm r} + K$. The dip also splits into two absorption branches, which corresponds to the transition from the dressed states composed of $\ket{0}$ and $\ket{1}$ to the less populated $\ket{2}$-like state.  
We determined $|K(0.31)|/2\pi$ to be 2.6~MHz~$\sim2 \kappa_{\rm tot}$ from the center frequency of the absorption just above the drive power, at which the $\omega_{12}$ transition becomes stronger than that of $\omega_{01}$ [vertical dashed line in Fig.~\ref{fig3}(c)].
The small discrepancy in $|K(0.31)|$ between the one-tone and two-tone spectroscopies is probably due to the small $K$, which is comparable to $\kappa_{\rm tot}$ at this flux bias, and to the weak but finite power dependence of the frequencies of the two-photon absorption and the $\omega_{12}$ transition.

Figures~\ref{fig3}(b) and \ref{fig3}(d) show the spectra at even higher
$f = 0.40$, where $\omega_{\rm r}/2\pi$, $\kappa_{\rm e}/2\pi$ and $\kappa_{\rm i}/2\pi$ are measured to be 10.015~GHz, 0.74~MHz and 0.72~MHz, respectively.
Reflecting the larger $K$, the dip for the two-photon absorption in one-tone spectroscopy [Fig.~\ref{fig3}(b)] and that for the $\ket{1}$-to-$\ket{2}$ transition in two-tone spectroscopy [Fig.~\ref{fig3}(d)] are more clearly separated from the dip at $\Delta \omega_{\rm p} =0$. 
From the spectra, $|K(0.40)|/2\pi$ 
is determined to be 11~MHz, which is roughly 8 times as large as $\kappa_{\rm tot}$  and well in the single-photon Kerr regime.

The observed one-tone spectra are very different from those at $f=0.00$ shown in Fig.~\ref{fig2}, reflecting the discreteness of the energy levels of the JPO.
It is worth mentioning that analogous variation from continuous to discrete spectra is observed in the ac Stark shift of a superconducting qubit in the weak- and strong-dispersive limits, although not in a single device~\cite{Schuster2005,Schuster2007}.

In Figs.~\ref{fig3}(a$^\prime$)-(d$^\prime$), 
we show the simulated spectra corresponding to Figs.~\ref{fig3}(a)-(d), respectively. In the calculation, the parameters $\omega_{\rm r}$, $\kappa_{\rm e}$, $\kappa_{\rm i}$ and $K$ are determined by the spectroscopy measurements. The microwave powers are estimated by the fitting at $f=0.00$ [Fig.~\ref{fig2}(b)] and the independently measured frequency dependence of the line attenuation (with the overall precision of 2~dB). Therefore, there are no adjustable parameters in the simulation. The simulated spectra reproduce the experimental results very well, confirming that the change of 
the spectroscopy data at different $f$ is caused by the change of the relation of $\kappa_{\rm tot}$ and $|K|$. 

\begin{figure}
\begin{center} 
\includegraphics[width=0.75\linewidth]{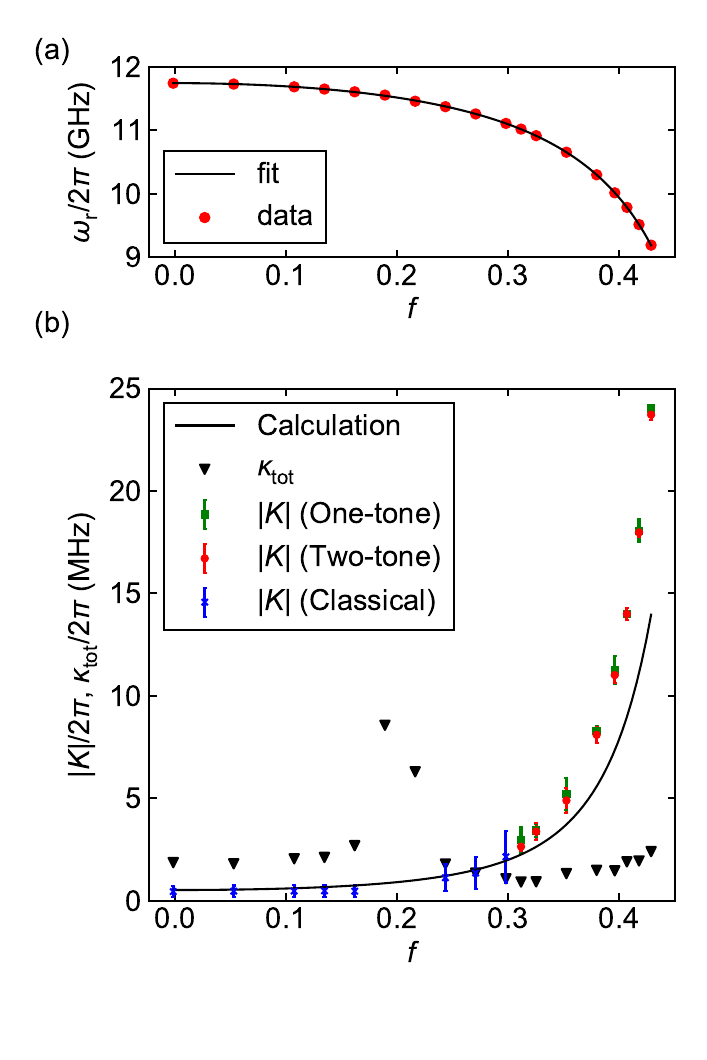}
\caption{
(a) Flux-bias dependence of the resonance frequency $\omega_{\rm r}$ of the JPO. 
Solid circles represent the measured $\omega_{\rm r}$, while the solid curve represents 
the theoretical calculation fitted to the data. 
(b) Flux-bias dependence of the Kerr nonlinearity $|K|$ and total photon loss rate $\kappa_{\rm tot}$. 
Blue crosses show $|K|$ estimated by the one-tone spectroscopy in the classical regime, while 
green squares and red circles respectively show $|K|$ measured by the one-tone and two-tone spectroscopies in the single-photon Kerr regime. The error bars of measured nonlinearities in the classical regime represent systematic uncertainties due to the power calibration (2~dB). The error bars in the single-photon Kerr regime are determined from the power dependence of the transition frequency of the two-photon absorption (one-tone) and $\omega_{12}$ (two-tone). The triangles show the measured $\kappa_\mathrm{tot}$.
Solid curve shows the theoretical calculation of $|K|$ using the critical current of Josephson junctions, 
which is determined by the fitting shown in (a). 
}
\label{fig4}
\end{center}
\end{figure}

\subsection{Flux-bias dependence of Kerr nonlinearity}
Figure~\ref{fig4}(a) shows the flux-bias dependence of $\omega_{\rm r}$. 
The measured $\omega_{\rm r}$ is well fitted by the theoretical calculation based on the equivalent-circuit model of a distributed-element resonator~\cite{Bourassa12}. 
In the calculation, we used the phase velocity of $0.398c$ ($c$ is the speed of light) and the characteristic impedance of 49.8~$\Omega$, 
which are determined from an independent measurement of a test CPW resonator. 
Regarding the SQUID as a linear inductor in parallel with a junction capacitance, 
we solve the transcendental equation for the distributed-element resonator 
to obtain the frequency of the fundamental resonance mode.
From the calculation, $I_{\rm c}$ of each Josephson junction in the SQUID 
is determined as a fitting parameter to be 0.93~$\mu$A, which is roughly consistent with the 
room-temperature resistance of a test junction on the same chip. As discussed below, the theoretical model also allows us to calculate the Kerr nonlinearity by taking into account the nonlinearity of the Josephson junctions at the lowest order~\cite{Bourassa12}.  


\begin{figure*}
\includegraphics[width=\linewidth]{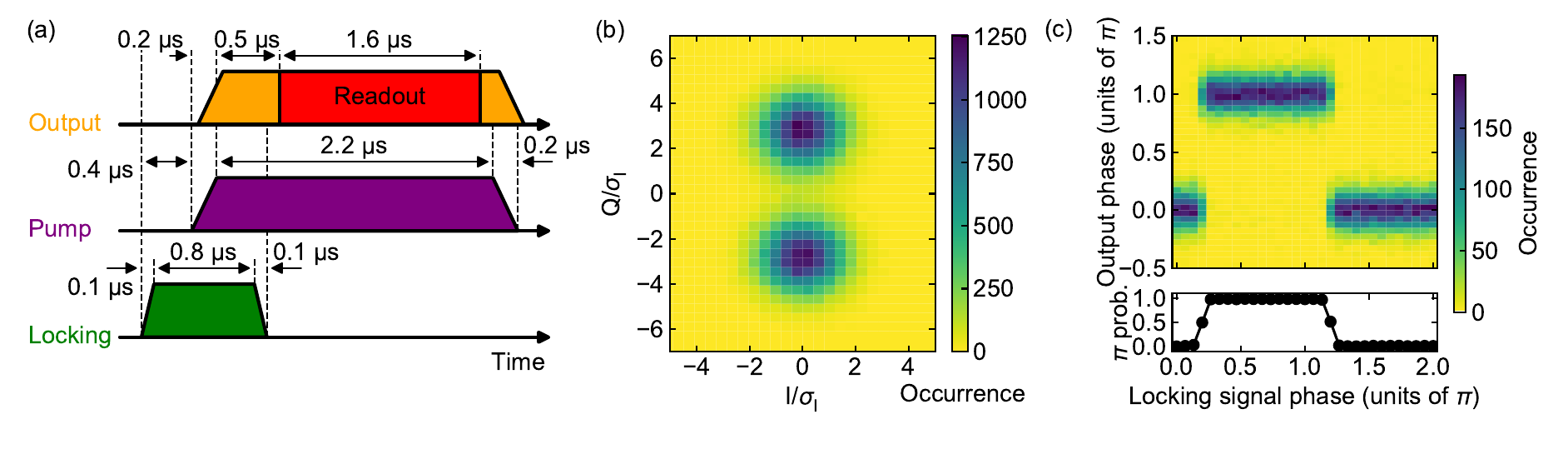}
\caption{Time-domain measurement at $f=0.40$. 
(a)~Pulse sequence. The pump power and the pump frequency are set to $-75$~dBm and $19.797$~GHz, respectively. The slope of pump and locking signal pulses are 0.2 and 0.1~${\rm \mu s}$, respectively.
(b)~Histogram of the output signal plotted in the in-phase and quadrature (IQ) plane without locking signals. The I and Q values are normalized by the standard deviation of the I values.
(c)~Histogram of the phase of the JPO output as a function of the locking signal phase relative to the pump (the upper graph). 
Power of the locking signal is $-89$~dBm. 
For each locking signal phase, $1\times 10^3$ pump pulses are applied and 
the probability of obtaining a $\pi$ state is calculated from the distribution (the lower graph). 
} 
\label{fig5}
\end{figure*}


Figure~\ref{fig4}(b) shows $|K|$ and $\kappa_{\rm tot}$ as a function of $f$. 
The loss rate $\kappa_{\rm tot}$ is obtained by fitting the reflection coefficient with the input--output theory in the low probe-power limit.
It is only weakly dependent on $f$, except around $f=0.2$, 
where the JPO is in resonance with a flux-independent spurious mode at 11.46~GHz. 
We attribute the mode to the chip mode. 
The Kerr nonlinearity $K$ is evaluated from the spectroscopy data. 
In the classical regime ($f\le 0.3$), $K$ is determined by fitting the measured probe-power dependence of $\omega_{\rm r}$ to the simulated one in the same way as in Fig.~\ref{fig2}.
The systematic uncertainties in the measured nonlinearity are given by the uncertainty in the power calibration of the probe tone (2~dB). 
In the single-photon Kerr regime ($f>0.3$), on the other hand, $K$ is determined from the frequency of the two-photon absorption dip in the one-tone spectroscopy 
or from the resonance at $\omega_{12}$ in the two-tone spectroscopy. The uncertainties in $K$ are evaluated by their dependence on the probe or drive powers, which are caused by Rabi oscillation.
The magnitudes of $K$ estimated by the one-tone and two-tone spectroscopy are consistent with each other within the systematic uncertainties.
As seen in the figure, $|K|$ is roughly equal to $\kappa_{\rm tot}$ at $f=0.27$, 
and is controllable over a wide range from smaller to larger than $\kappa_{\rm tot}$ in a single device. 
This allows us to observe the change of spectra between the classical and single-photon Kerr regimes. 

The solid curve in Fig.~\ref{fig4}(b) represents the theoretical prediction of the Kerr nonlinearity using the critical current of the SQUID, which is obtained in the fitting of Fig.~\ref{fig4}(a). 
It reproduces $K$ reasonably well as a function of $f$ in the small $f$ region ($f<0.35$) within the uncertainties of $|K|$. Small systematic deviation between theory and the experiment is possibly due to our choice of the initial value in the fitting. 

At higher $f$, experimentally obtained $|K|$ becomes systematically higher than the theoretical prediction, possibly because the approximation of the theoretical model, which considers only the lowest nonlinearity, is not valid in the region $f\sim 0.5$ where the effective critical current of the SQUID becomes smaller.

\section{parametric oscillation in the single-photon Kerr regime}
In this section we investigate the property of parametric oscillations in the single-photon Kerr regime at $f$ = 0.40, 
which is important for applications of JPO in quantum information processing such as quantum annealing~\cite{Goto2016_scirep}. 
In order to fix the pump frequency and amplitude for the time-domain measurements, we first measured the dependence of the parametric oscillation on the frequency $\omega_\mathrm{m}$ and pump power $P_\mathrm{m}$ by using a continuous pump microwave and a spectrum analyzer. We observed the highest output power of the parametric oscillation, $P_{\rm o}=-122$~dBm, which corresponds to the mean photon number in the resonator of $n_{\rm o} = P_{\rm o}/(\hbar \omega_{\rm r}\kappa_{\rm e})=20$ at $\omega_{\rm m}/2\pi = 19.797$~GHz and $P_{\rm m}=-75$~dBm. The half of the pump frequency is negatively detuned from $\omega_{\rm r}/2\pi$ by $\sim120$~MHz possibly due to pump-induced shift of resonance frequency. Further investigation is required to clarify the cause of the large detuning quantitatively.  

By fixing the amplitude and the frequency of the pump in this way, we performed the time-domain measurement using a pulsed pump. 
A pulsed pump tone is applied, and the output signal is recorded in the heterodyne measurement with the pulse sequence shown in Fig.~\ref{fig5}(a) without a locking signal.
Figure~\ref{fig5}(b) shows the histogram of the output signal plotted in the in-phase and quadrature (IQ) plane. 
We observe two equally distributed peaks with an equal amplitude and different phases shifted by $\pi$, 
which is the indication of parametric oscillations~\cite{Wilson2010, Krantz2013, Wustmann2013}. 

We further checked the controllability of the distribution of these states by applying an external locking signal~\cite{Lin2014}. 
The locking signal with half a pump frequency and a power of $P_{\rm s}=-89$~dBm at the device 
is injected from the feed line of the JPO and is turned off during the pump.
The output signal from the JPO is integrated after the locking signal is turned off [Fig.~\ref{fig5}(a)]. 
Figure~\ref{fig5}(c) shows the histogram of the phase of the JPO output as a function of the relative phase between the locking signal and the pump.
It is clearly seen that the probability distribution of the output states depends on the relative phase, 
indicating that the probability of having $0$ or $\pi$ state can be controlled by the phase of the locking signal.
The locking error is $\sim1\%$, which we attribute to the switching of the output state during the readout. As the input signal power decreases, the locking error increases (See Appendix F). The input signal power of $-89$ dBm required to achieve the minimum phase lock error of 1\% corresponds to the mean photon number of $4P_{\rm s}\kappa_{\rm e}/(\hbar \omega_{\rm r}\kappa^2) = 4\times 10^4$, which is $\mathcal{O}(10^4)$ times higher than the previous study using a JPO in the classical regime \cite{Lin2014}. 
It is probably due to the large nonlinearity of the JPO and the large oscillation amplitude at the operation point of the time-domain measurement, which result in a large potential barrier between the oscillation states $\ket{\pm \alpha}$. The locking signal manipulates the probability distribution of the oscillation states by tilting the metapotential of JPO, and the signal strength required for this manipulation depends on the height of the potential barrier. The locking signal strength required to overwhelm the potential barrier is proportional to $(K/\kappa)^2 \alpha^6$ (see Appendix~E for details). In the present experiment, the estimated value is $\mathcal{O}\left(10^4 \right)$ times higher than that of the previous study \cite{Lin2014}, which is consistent with the experiment. 
The estimated value is also $\sim 600$ higher than that of the KPO studied in Ref.~\citenum{Wang2019} due to the large oscillation amplitude $\alpha$, which indicates that the oscillation amplitude should be weakened to reduce the required locking signal strength.

\section{conclusion}
We have performed spectroscopy measurements in a JPO with a controllable Kerr nonlinearity, which cover both classical and single-photon Kerr regimes of a nonlinear oscillator.
The measured spectra show a distinct difference between the two regimes and are well reproduced by numerical simulations.
In the single-photon Kerr regime, we have also observed parametric oscillations with a well-defined phase of either $0$ or $\pi$, 
whose probability can be controlled with an externally injected signal. The signal strength required to achieve the phase locking is orders of magnitude higher than that of previous studies.

The present work is an important step towards applications of JPO in quantum information processing.
Although we focused on the spectroscopic behavior in the crossover regime here, other properties such as the stability of the parametric oscillation \cite{Wustmann2017, Sivak2019} and the switching rate between the two oscillation states \cite{Marthaler2006, Andersen2020, Muppalla2018} in the crossover regime are also important and relatively unexplored. The present system provides a platform to study those properties, which will be an interesting topic for the future study. 

\section{ACKNOWLEDGMENTS}
We thank Y. Urade and K. Zuo for their help in experiments at the early stage of this work. 
We also thank A. Noguchi and K. Inomata for fruitful discussions, 
Y. Hashimoto for the EM simulation, and A. Morioka and Y. Kitagawa for their assistance in the device fabrication.
This work was partly supported by JST ERATO (Grant No.~JPMJER1601). 
This paper is based on results obtained from a project,
JPNP16007, commissioned by the New Energy and Industrial Technology Development
Organization (NEDO). 

\appendix
\counterwithin{figure}{section}
\renewcommand\thefigure{\thesection\arabic{figure}}
\section{Device details}
\begin{figure}
\begin{center} 
\includegraphics[width=\linewidth]{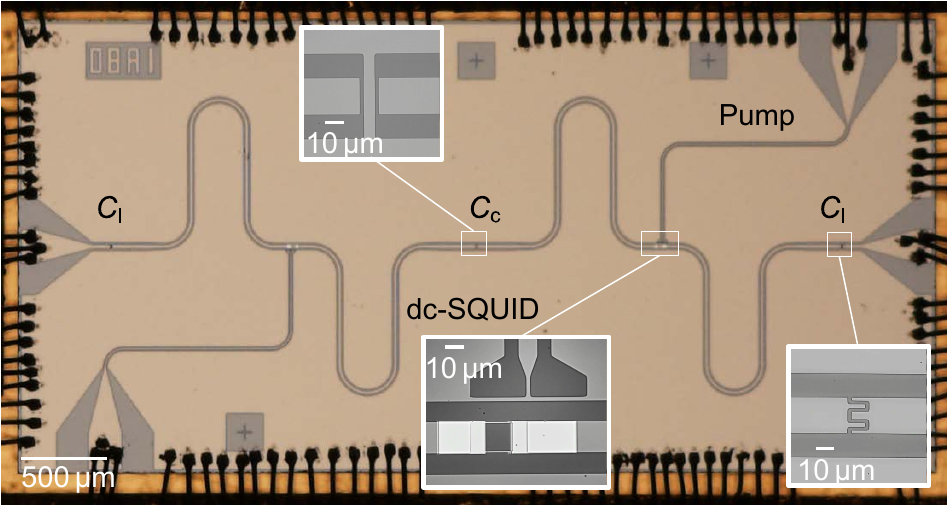}
\caption{Optical image of the JPO chip. 
The chip contains two JPOs, and the JPO on the right-hand side is studied in the paper. 
The insets show magnified views around the SQUID and capacitors.} 
\label{fig6}
\end{center}
\end{figure}

Figure~\ref{fig6} shows an optical image of the chip containing the JPO studied in this paper. 
The coplanar waveguide resonator is etched from a 100-nm thick Nb film sputtered 
on a 380-$\mu$m thick high-resistivity Si substrate, 
and the SQUID is fabricated by shadow evaporation of Al after cleaning the Nb surface by Ar ion milling. 

The chip contains two JPOs which are coupled by a {0.6-fF} capacitor denoted as $C_{\rm c}$ in the center of the picture. As seen in the inset for $C_{\rm c}$, each end of the resonators is also connected to the ground by a capacitance ($C_{\rm I}^\prime$ in the main text), which is estimated to be 3.3 fF. 
We studied the JPO on the right-hand side in this paper, while keeping the left-hand side JPO far detuned.

\section{Dilution refrigerator setup}
 \begin{figure}
	\includegraphics[width=\linewidth]{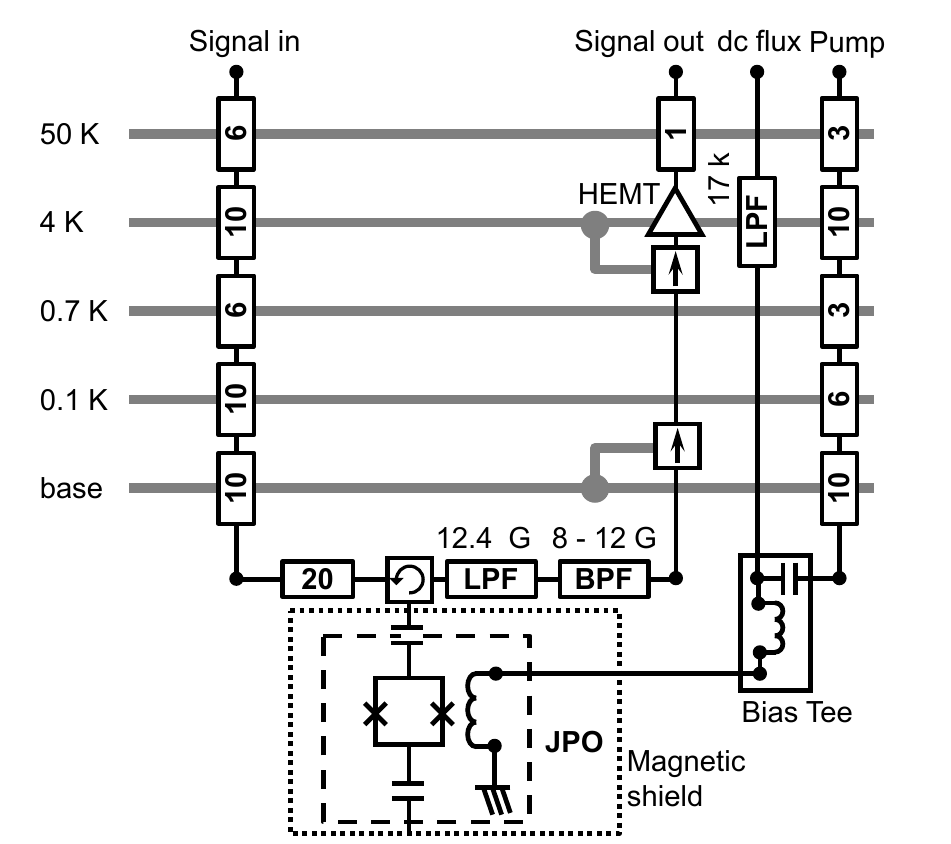}
	\caption{Schematic of the setup in the dilution refrigerator. The horizontal gray lines show the thermal stages of the fridge. The rectangles with a number inside represent fixed attenuators with corresponding attenuation in units of dB. The rectangles with a circular and straight arrows inside represent a circulator and an isolator, respectively. LPF and BPF mean a low-pass and a band-pass filters, respectively.
	 }
	\label{fig:rf_fridge}
\end{figure}

Figure~\ref{fig:rf_fridge} shows the schematic of the setup in the dilution refrigerator. The sample is installed inside a magnetic shield and thermally anchored to the mixing chamber with a base temperature less than 10~mK. The input and pump lines are $50-\Omega$ rf coaxial cables equipped with attenuators thermally anchored to each stage of the dilution refrigerator to reduce the thermal noise from the room-temperature environment. 
The probe microwave is injected into the sample via a circulator to separate the input and output. The output is filtered with a low-pass filter (LPF) and a bandpass filter (BPF), and then amplified with a cryogenic high electric mobility transistor (HEMT) amplifier.
The pump line is combined with the DC current line by a bias tee, which is connected to the pump port of the JPO.

\section{Theory for the simulation of spectroscopy experiments}

\begin{figure}
  \begin{center} 
    \includegraphics[width=\linewidth]{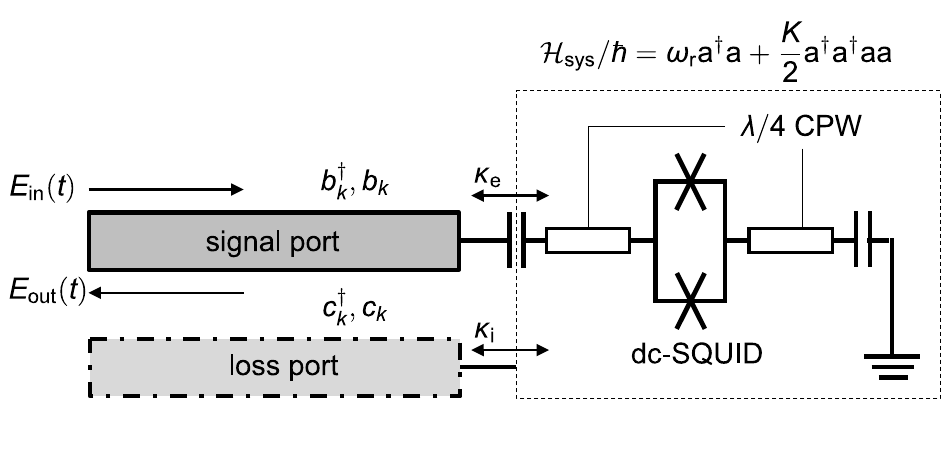}
    \caption{Schematic diagram of a nonlinear-resonator model. The resonator with the resonance frequency $\omega_{\rm r}$ and the Kerr nonlinearity $K$ is coupled to the signal and fictitious loss ports with the external and internal photon loss rates $\kappa_{\rm e}$ and $\kappa_{\rm i}$, respectively. The input and output fields with their respective amplitudes of $E_{\rm in}(t)$ and $E_{\rm out}(t)$ propagate through the signal port. The operators $a$, $b_k$ and $c_k$ represent the annihilation operators of the fundamental mode of the resonator and plain-wave modes of the two ports, respectively.} 
    \label{fig7}
  \end{center}
\end{figure}

In this section, we discuss the theoretical model to simulate the one-tone and two-tone spectroscopy experiments shown in the main text. 
We consider a model presented in Fig.~\ref{fig7}. A nonlinear resonator with the resonance frequency of $\omega_{\rm r}$ 
and the Kerr nonlinearity of $K$ is probed by an external signal via the signal port, which is a semi-infinite waveguide coupled to the resonator with the coupling (or decay) rate of $\kappa_{\rm e}$.
We denote the amplitudes of the input and output signals by $E_{\rm in}(t)$ and $E_{\rm out}(t)$, respectively.
The fictitious loss port coupled to the resonator with the coupling rate of $\kappa_{\rm i}$ is assumed additionally to model the internal loss.
The loss rates can be written as $\kappa_{\rm e}=\omega_{\rm r}/Q_{\rm e}$ and $\kappa_{\rm i}=\omega_{\rm r}/Q_{\rm i}$, respectively, 
where $Q_{\rm e}$ and $Q_{\rm i}$ are the external and internal quality factors of the resonator, respectively.

The Hamiltonian of the system can be represented as follows,
\begin{eqnarray}
\mathcal{H}&=&\mathcal{H}_{\rm sys}+\mathcal{H}_{\rm sig}+\mathcal{H}_{\rm loss}, \\
\mathcal{H}_{\rm sys}/\hbar&=&\omega_{\rm r} a^\dag a + \frac{K}{2}a^\dag a^\dag a a, \label{eq:sysh} \\
\mathcal{H}_{\rm sig}/\hbar&=&\nonumber \\
\int &dk& \left[v_b k b_k^\dag b_k +\sqrt{\frac{v_b \kappa_{\rm e}}{2\pi}}\left(a^\dag b_k + b^\dag_k a\right)\right], \\
\mathcal{H}_{\rm loss}/\hbar&=&\nonumber\\
\int &dk& \left[v_c k c_k^\dag c_k +\sqrt{\frac{v_c \kappa_{\rm i}}{2\pi}}\left(a^\dag c_k + c^\dag_k a\right)\right],
\end{eqnarray}
where $a$ is the annihilation operator of the fundamental mode of the resonator, $b_k$ and $c_k$ are the annihilation operators of plane-wave modes of the signal and fictitious loss ports with a wavenumber of $k$, respectively, and $v_{b}$ and $v_{c}$ are the phase velocities of the corresponding ports. 
The annihilation operators have the following commutation relations, $[a,a^\dag]=1$, $[b_k, b_{k^\prime}^\dag]=\delta(k-k^\prime)$ and $[c_k, c_{k^\prime}^\dag]=\delta(k-k^\prime)$. 

From the Heisenberg equations of motion for $b_k$, we obtain
\begin{eqnarray}
\frac{db_k(t)}{dt} &=& \frac{\mathrm{i}}{\hbar}[\mathcal{H}, b_k(t)]\nonumber \\
&=&-\mathrm{i}v_b k b_k(t) -\mathrm{i} \sqrt{\frac{v_b\kappa_{\rm e}}{2\pi}}\, a(t).
\end{eqnarray}
The equation can be solved formally as follows,
\begin{eqnarray}
b_k(t)&=&e^{-\mathrm{i}v_b k t}b_k(0)\nonumber \\
&&-\mathrm{i}\sqrt{\frac{v_b\kappa_{\rm e}}{2\pi}}\int_0^t dt^\prime e^{-\mathrm{i}v_b k(t-t^\prime)}a(t^\prime). \label{eq:formal}
\end{eqnarray}
In order to simplify the formal solution, we introduce the real-space representation of the waveguide field 
$\tilde{b}_r\equiv\frac{1}{\sqrt{2\pi}}\int dk e^{\mathrm{i}kr}b_k$, 
where operators with $r<0$ and $r>0$ correspond to the incoming and outgoing fields, respectively. 
The simultaneous operators with different degrees of freedom commute with each other. For example, 
\begin{eqnarray}
[\tilde{b}_r(t), a^\dag(t)] = [\tilde{c}_r(t), a^\dag(t)] = 0.
\end{eqnarray}
By using this representation, Eq.~(\ref{eq:formal}) becomes
\begin{eqnarray}
\tilde{b}_r(t) &=& \tilde{b}_{r-v_b t}(0)\nonumber\\
&&-\mathrm{i}\sqrt{\frac{\kappa_{\rm e}}{v_b}}\,\theta(r)\,\theta\!\left(t-\frac{r}{v_b}\right)a\!\left(t-\frac{r}{v_b}\right),
\end{eqnarray}
where the second term with the Heviside step function $\theta(r)$ is non-zero under the condition that $r$ is within the range of $[0, v_b t]$. This equation implies that the output field is a superposition of the input field and the response from the resonator. 
A corresponding equation for the loss port is obtained in the same way: 
\begin{eqnarray}
\tilde{c}_r(t) &=& \tilde{c}_{r-v_c t}(0)\nonumber\\
&&-\mathrm{i}\sqrt{\frac{\kappa_{\rm i}}{v_c}}\,\theta(r)\,\theta\!\left(t-\frac{r}{v_c}\right)a\!\left(t-\frac{r}{v_c}\right).
\end{eqnarray}

For simplicity, we set phase velocities to unity in the following discussion. 
The Hamiltonian related to the resonator mode (omitting terms proportional to $b_k^\dagger b_k$ or $c_k^\dagger c_k$) and the input-output relation can be written in this case as follows, 
\begin{eqnarray}
\mathcal{H}_a/\hbar &=& \omega_{\rm r} a^\dag a +\frac{K}{2}a^\dag a^\dag a a\nonumber\\
&&+ \sqrt{\kappa_{\rm e}}\,(a^\dag \tilde{b}_0 + \tilde{b}_0^\dag a)\nonumber\\
&& + \sqrt{\kappa_{\rm i}}\,(a^\dag \tilde{c}_0 +\tilde{c}_0^\dag a), \label{eq:ha}\\
\tilde{b}_r(t) &=&\tilde{b}_{r-t}(0) - \mathrm{i}\sqrt{\kappa_{\rm e}}\,\theta(r)\,\theta\!\left(t-r \right)a\!\left(t-r\right), \label{eq:b-a}\\
\tilde{c}_r(t) &=&\tilde{c}_{r-t}(0) - \mathrm{i}\sqrt{\kappa_{\rm i}}\,\theta(r)\,\theta\!\left(t-r \right)a\!\left(t-r\right). \label{eq:c-a}
\end{eqnarray}
The fields at $r=0,\ t>0$ in Eq.~(\ref{eq:ha}) are reduced to $\tilde{b}_0(t)=\tilde{b}_{-t}(0)-i\frac{\sqrt{\kappa_{\rm e}}}{2}\,a(t)$ and $\tilde{c}_0(t)=\tilde{c}_{-t}(0)-i\frac{\sqrt{\kappa_{\rm i}}}{2}\,a(t)$, where we use the definition of the Heaviside step function, $\theta(0)=1/2$.
We define operators corresponding to the signal injection from the signal and loss ports as $b_{\rm in}(t) \equiv \widetilde{b}_{-0}(t)=\widetilde{b}_{-t}(0)$ and $c_{\rm in}(t) \equiv \widetilde{c}_{-0}(t)=\widetilde{c}_{-t}(0)$, respectively. The operator corresponding to the output signal can be defined in the same way, 
\begin{equation}
b_{\rm out}(t) \equiv \widetilde{b}_{+0}(t)=\widetilde{b}_{-t}(0)-i\sqrt{\kappa_{\rm e}}\,a(t). \label{eq:bout}
\end{equation}

The microwave response of the resonator can be described by the input--output relation and the operators $A_{m,n}\equiv (a^\dag)^m a^n$, where $m$ and $n$ are non-negative integers.
The Heisenberg equation for this operator is given by
\begin{widetext}
\begin{eqnarray}
-{\rm i} \frac{dA_{m,n}}{dt}&=&\left[(m-n)\omega_{\rm r} + \frac{(m-n)(m+n-1)K}{2}+{\rm i}\frac{(m+n)\kappa_{\rm tot}}{2}\right]A_{m,n}+ (m-n)KA_{m+1,n+1}\nonumber \\
&&- nA_{m,n-1}[\sqrt{\kappa_{\rm e}}\,{b}_{\rm in}(t)+\sqrt{\kappa_{\rm i}}\,{c}_{\rm in}(t)]+ m[\sqrt{\kappa_{\rm e}}\,{b}_{\rm in}^\dag(t)+\sqrt{\kappa_{\rm i}}\,{c}_{\rm in}^\dag(t)]A_{m-1,n}, \label{eq:A_matrix}
\end{eqnarray}
\end{widetext}
where $\kappa_{\rm tot}\equiv \kappa_{\rm e}+\kappa_{\rm i}$.

Now we consider the microwave response to two-tone inputs by neglecting the injection from the loss port. The mean value of $b_{\rm in}$ and $c_{\rm in}$ can be represented as
\begin{eqnarray}
\langle b_{\rm in}(t)\rangle &=& E_{\rm p} e^{-\mathrm{i}\omega_{\rm p} t} + E_{\rm d} e^{-\mathrm{i}\omega_{\rm d} t},\label{eq:bin}\\
\langle c_{\rm in}(t)\rangle &=& 0,\label{eq:cin}
\end{eqnarray}
where $E_{\rm p(d)}$ and $\omega_{\rm p(d)}$ are the amplitude and frequency of the probe (drive) tone, respectively. The amplitude $E_{\rm p(d)}$ is related to the input power $P_{\rm p(d)}$ by
\begin{equation}
E_{\rm p(d)}=\sqrt{\frac{P_{\rm p(d)}}{\hbar \omega_{\rm p(d)}}}.
\end{equation}
We switch to the frame rotating at the probe ($i$=p) or drive ($i$=d) frequency by
\begin{eqnarray}
\langle A_{m,n}\rangle=\alpha_{m,n}(t) e^{\mathrm{i}(m-n)\omega_{i} t}.\label{eq:amn}
\end{eqnarray}
By taking the mean value and using Eqs.~(\ref{eq:bin}) to (\ref{eq:amn}), Eq.~(\ref{eq:A_matrix}) can be reduced to 
\begin{widetext}
\begin{eqnarray}
-{\rm i}\frac{d}{d t}\alpha_{m,n} &=&\left[(m-n)(\omega_{\rm r} - \omega_i) +\frac{(m-n)(m+n-1)K}{2}+{\rm i} \frac{(m+n)\kappa_{\rm tot}}{2}\right]\alpha_{m,n}
+(m-n)K\alpha_{m+1,n+1}\nonumber \\
&&-n\sqrt{\kappa_{\rm e}}\,\alpha_{m,n-1}\left(E_{\rm p}e^{\mathrm{i}(\omega_{i}-\omega_{\rm p})t}+E_{\rm d}e^{\mathrm{i}(\omega_{i}-\omega_{\rm d})t}\right)\nonumber \\
&&+ m\sqrt{\kappa_{\rm e}}\,\alpha_{m-1,n}\left(E_{\rm p}^\ast e^{-\mathrm{i}(\omega_{i}-\omega_{\rm p})t} +E_{\rm d}^\ast e^{-\mathrm{i}(\omega_{i}-\omega_{\rm d})t}\right). \label{eq:master}
\end{eqnarray}
\end{widetext}
The reflection coefficient $\Gamma$ of the probe microwave can be obtained by solving the equation and using the input-output relation [the mean value of Eq.~(\ref{eq:bout})],
\begin{eqnarray}
E_{\rm out}(t)&=&E_{\rm in}(t)-\mathrm{i}\sqrt{\kappa_{\rm e}}\langle A_{0,1}(t)\rangle, \\
\Gamma &=& 1-\mathrm{i}\sqrt{\kappa_{\rm e}}\,\frac{\alpha_{0,1}^{\rm p}}{E_{\rm p}}. \label{eq:spara}
\end{eqnarray}
In the following subsections, we consider the two cases of the one-tone and two-tone spectroscopies.

\subsection{One-tone spectroscopy}
When the input is one-tone, namely $E_{\rm d}=0$, the system is static in the rotating frame with the probe frequency ($i={\rm p}$). Accordingly, we set $\alpha_{m,n}(t)=\alpha_{m,n}^{\rm p}$, where $\alpha_{m,n}^{\rm p}$ is time-independent. Thus, Eq.~(\ref{eq:master}) is reduced to

\begin{widetext} 
\begin{eqnarray}
0 &=&\left[(m-n)(\omega_{\rm r} - \omega_{\rm p})+\frac{(m-n)(m+n-1)K}{2}+{\rm i}\frac{(m+n)\kappa_{\rm tot}}{2}\right]\alpha_{m,n}^{\rm p} \nonumber \\
&&+K(m-n)\alpha_{m+1,n+1}^{\rm p} -n\sqrt{\kappa_{\rm e}}\,E_{\rm p}\alpha_{m,n-1}^{\rm p} + m\sqrt{\kappa_{\rm e}}\,E_{\rm p}^\ast \alpha_{m-1,n}^{\rm p}. \label{eq:one-tone}
\end{eqnarray}
\end{widetext}

\subsection{Two-tone spectroscopy}
The linear response to the probe wave can be deduced from Eq.~(\ref{eq:master}) in the rotating frame with the drive frequency ($i={\rm d}$). In this case, $\alpha_{m,n}(t)$ can be written as
\begin{eqnarray}
\alpha_{m,n}(t)=\alpha_{m,n}^{\rm d}+e^{\mathrm{i}(\omega_{\rm d}-\omega_{\rm p})t}\alpha_{m,n}^{\rm p}.
\end{eqnarray}
We solve the equations for $\alpha_{\rm m, n}^{\rm d}$ and $\alpha_{\rm m, n}^{\rm p}$ perturbatively with respect to the probe amplitude. Firstly, we assume the probe tone is much weaker than the drive tone and solve Eq.~(\ref{eq:master}) for $\alpha_{\rm m, n}^{\rm d}$ by neglecting the contribution from the probe input ($\alpha_{\rm m, n}^{\rm p}=0$).
The equation for $\alpha_{\rm m, n}^{\rm d}$ is the same as Eq.~(\ref{eq:one-tone}) except that $\omega_{\rm p}$, $\alpha_{m, n}^{\rm p}$ and $E_{\rm p}$ are replaced with $\omega_{\rm d}$, $\alpha_{m, n}^{\rm d}$ and $E_{\rm d}$, respectively.
Secondly, we solve Eq.~(\ref{eq:master}) for $\alpha_{\rm m, n}^{\rm p}$ by extracting factors proportional to $e^{\mathrm{i}(\omega_{\rm d}-\omega_{\rm p})t}$ and using $\alpha_{\rm m, n}^{\rm d}$ obtained in the first step, namely,
\begin{widetext}
\begin{eqnarray}
n \sqrt{\kappa_{\rm e}}\,E_{\rm p} \alpha_{m,n-1}^{d} &=&\left[(m-n)(\omega_{\rm r} - \omega_d) -(\omega_d-\omega_p)+ \frac{(m-n)(m+n-1)K}{2}
+{\rm i}\frac{(m+n)\kappa_{\rm tot}}{2}\right]\alpha_{m,n}^{\rm p} \nonumber \\
&&+(m-n)K\alpha_{m+1,n+1}^{\rm p} -n\sqrt{\kappa_{\rm e}}\,E_{\rm d}\alpha_{m,n-1}^{\rm p} + m\sqrt{\kappa_{\rm e}}\,E_{\rm d}^\ast \alpha_{m-1,n}^{\rm p}. \label{eq:two-tone}
\end{eqnarray}
\end{widetext}

\section{Rabi splitting}

\begin{figure}
  \begin{center} 
    \includegraphics[width=\linewidth]{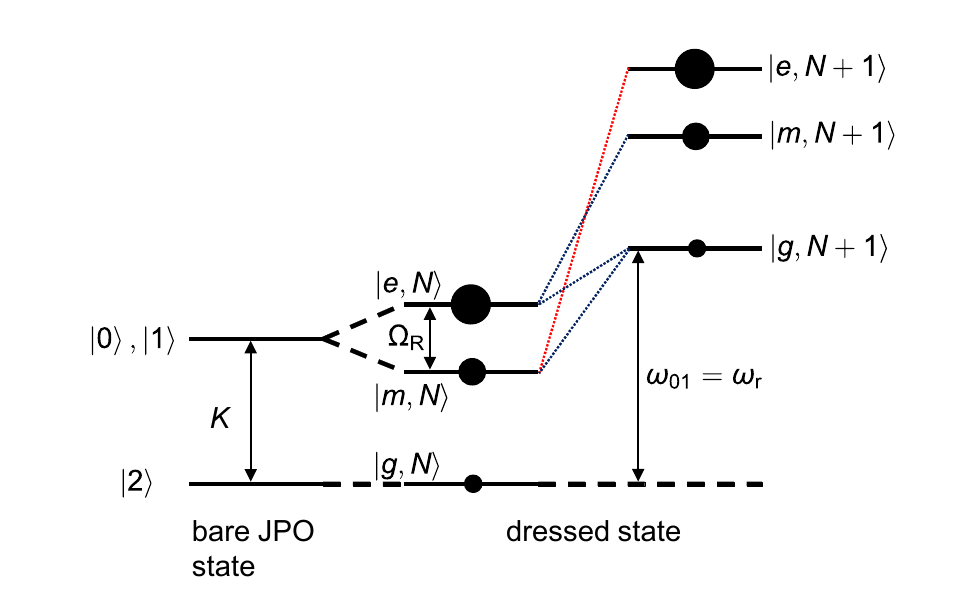}
    \caption{Rabi splitting in two-tone spectroscopy. Left: the energy diagram of the lowest three Fock states in a frame rotating at a drive frequency, $\omega_{\rm d}=\omega_{01}$. Center and Right: the energy diagram of the dressed states with a photon number of $N$ and $N+1$, respectively. The red and blue dashed lines show transitions that gives amplification and attenuation of the probe, respectively.} 
    \label{fig:rabi}
  \end{center}
\end{figure}

The strong drive (probe) tone in two-tone (one-tone) spectroscopy induces Rabi oscillations between the Fock states of the resonator, which causes systematic uncertainties of the estimated Kerr nonlinearities. Figure~\ref{fig:rabi} shows the Rabi splitting in the two-tone spectroscopy \cite{Koshino2013}. The three lowest Fock states of the resonator, $\ket{0}$, $\ket{1}$ and $\ket{2}$, in the rotating frame at a drive frequency of $\omega_{01}=\omega_{\rm r}$ are shown. The lowest two Fock states, $\ket{0}$ and $\ket{1}$, are degenerate in the frame and the state $\ket{2}$ has an energy shifted by $K$ (negative for JPOs). The strong drive tone induces Rabi oscillations between the degenerate Fock states, which are then mixed into two dressed states with an energy difference of Rabi frequency, $\Omega_{\rm R}\equiv \sqrt{\kappa_{\rm e}}E_{\rm d}$. The $\ket{2}$ state is not affected by the drive tone when the magnitude of Kerr nonlinearity is much larger than $\Omega_{\rm R}$. We label the lowest three levels under the strong drive tone as $\ket{e, N}$, $\ket{m, N}$ and $\ket{g, N}$, where $N$ represents the photon number.
As shown in Ref.~\citenum{Koshino2013}, the population of $\ket{e, N}$ state becomes larger than that of $\ket{m, N}$ state under the strong drive. The population of $\ket{g, N}$ state is the lowest as it is close to the $\ket{2}$ state. 

The resonance in two-tone spectroscopy can be understood as the transition between dressed states with the photon number of $N$ and $N+1$, whose energy difference is the drive frequency $\omega_{\rm d}=\omega_{\rm 01}$. The transitions between $\ket{e, N+1}$ and $\ket{m, N}$ and between $\ket{m, N+1}$ and $\ket{e, N}$ correspond to the fundamental transition at $\omega_{01}$ at weak drive power limit, and the energy splitting between these transitions is $2\Omega_{\rm R}$. On the other hand, the transitions between $\ket{g, N+1}$ and $\ket{e, N}$ and between $\ket{g, N+1}$ and $\ket{m, N}$ correspond to the second transition at $\omega_{12}=\omega_{01}+K$, and the energy splitting between these states is $\Omega_{\rm R}$. The transition between $\ket{e, N+1}$ and $\ket{m, N}$ amplifies the probe tone because the higher energy state $\ket{e, N+1}$ has a population larger than $\ket{m, N}$. The other three transitions are absorptive because the lower energy state has larger population. Although the transitions between the dressed states with the same suffixes (such as the transition from $\ket{g, N+1}$ to $\ket{g, N}$) are not forbidden, their effects are not visible in the spectrum because the contribution of emission and absorption are canceled out \cite{Koshino2013}.

In one-tone spectroscopy, strong probe tone induces Rabi oscillation between $\ket{0}$ and $\ket{1}$ states and between $\ket{1}$ and $\ket{2}$ states, which shift the effective frequency of two-photon absorption \cite{Koshino2013}.

\section{Signal strength required for phase locking}
In this section, we estimate the signal strength required to fully lock the oscillation states of JPOs by considering their metapotential.
The dimensionless Hamiltonian of JPOs under phase locking favoring the $\ket{+\alpha}$ state can be written as follows~\cite{Yamamoto_book}, 
\begin{eqnarray}
\mathcal{H}^\prime&=&\mathcal{H}/(\hbar\kappa/2)\nonumber \\
&=&\gamma^\prime a^\dag a^\dag a a + \Delta^\prime a^\dag a +  \frac{\epsilon^\prime}{2} \left(a^\dag a^\dag + a a\right) \nonumber\\
&+&\sqrt{N_{\rm s}} \left(a^\dag + a \right),
\end{eqnarray}
where we use the rotating frame at the oscillation frequency $\omega_{\rm p}/2$, $\Delta^\prime\equiv 2(\omega_{\rm r} - \omega_{\rm p} / 2) / \kappa$ is the reduced detuning, $\gamma^\prime \equiv K/\kappa < 0$ is the dimensionless nonlinearity, $\epsilon^\prime$ is the pump amplitude normalized by the oscillation threshold, at which the modulation amplitude of the resonance frequency is equal to the loss rate $\kappa$, and $N_{\rm s}\equiv 4\kappa_{\rm e}\left|E_{\rm s}\right|^2 / \kappa^2$ is the strength of the locking signal. The signal strength corresponds to the mean photon number in the resonator in the case of $K=0$.
By replacing the creation and annihilation operators with a $c$-number $x$, we can deduce the following metapotential of JPOs, 
\begin{eqnarray}
U / (\hbar\kappa / 2) &=&\gamma^\prime \left|x\right|^4+\Delta^\prime |x|^2 + \frac{\epsilon^\prime}{2}\left(x^{\ast 2} + x^2\right) \label{eq-meta}\nonumber\\
&+& \sqrt{N_{\rm s}}\left(x^\ast + x\right).
\end{eqnarray}
The stationary points of the metapotential are located on the real axis in this case, and their positions correspond to the amplitude of the oscillation states when we neglect the single-photon loss. Although the single-photon loss makes the stationary points complex and lowers the oscillation amplitude \cite{Puri2017}, this effect is negligible when $\epsilon^\prime + \Delta^\prime$ is much larger than unity. When $N_{\rm s}=0$ and we neglect the effect of the single-photon loss, the oscillation amplitude is $\alpha= \sqrt{(\epsilon^\prime+\Delta^\prime) / (-2\gamma^\prime)}$ \cite{Goto2016_scirep}. The detuning term behaves similarly to the pump term in the metapotential, and the oscillation amplitude increases when the oscillation frequency is negatively detuned from the resonance frequency ($\Delta^\prime > 0$).
 
\begin{figure}
  \begin{center} 
    \includegraphics[width=\linewidth]{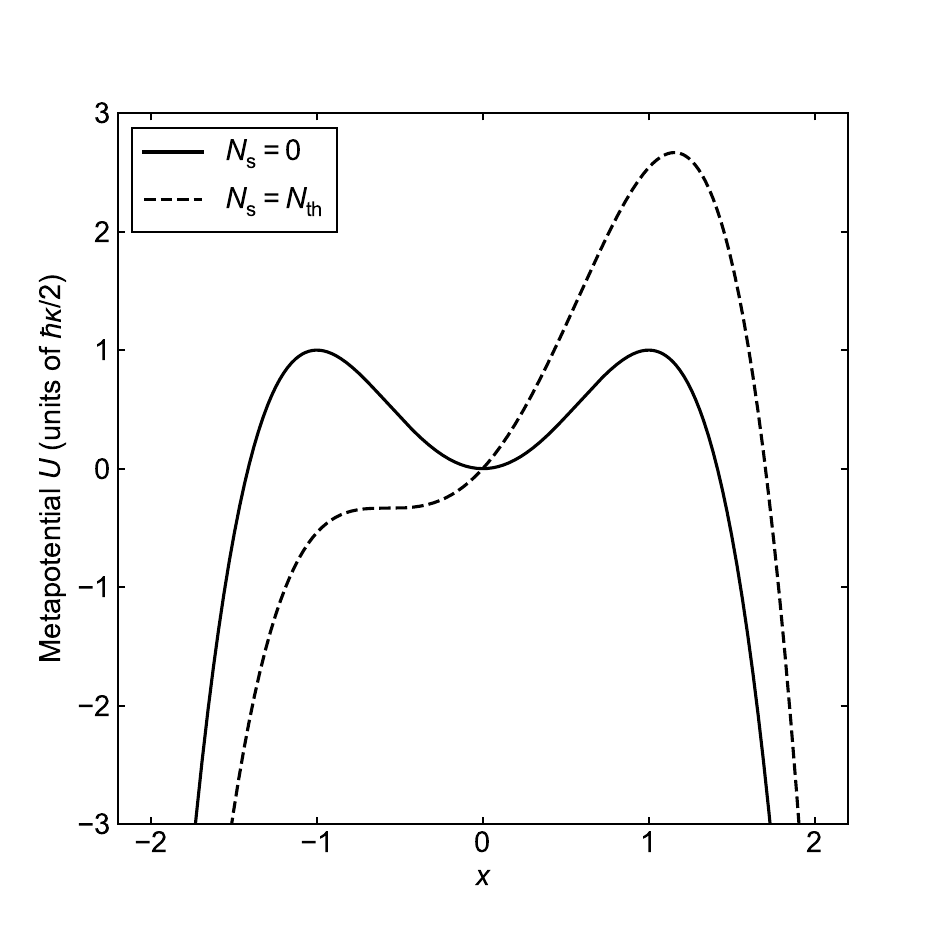}
    \caption{Metapotential under phase locking. The horizontal and vertical axes show the real amplitude $x$ and the normalized metapotential shown in Eq.~(\ref{eq-meta}). The solid and dashed lines show the metapotential at $N_{\rm s}=0$ and $N_{\rm s}=N_{\rm th}=16/27$, respectively, where the other parameters are $\gamma^\prime=-1$, $\epsilon^\prime+\Delta^\prime=2$ and $\alpha=1$.} 
    \label{fig:meta}
  \end{center}
\end{figure}

In the case of $N_{\rm s}$=0, the stationary points are identical to $\pm \alpha$ as shown by the solid line in Fig.~\ref{fig:meta}. The two stationary points at $x=\pm \alpha$ are the tops of the inverse double well potential. By imposing the phase locking signal, the metapotential is tilted and one of the stationary points $\ket{-\alpha}$ becomes unstable. The signal strength required to turn the potential into a single well potential is given by,
\begin{eqnarray}
N_{\rm th} = \frac{16}{27}\gamma^{\prime 2} \alpha^6.
\end{eqnarray}
The dashed line in Fig.~\ref{fig:meta} shows the metapotential at $N_{\rm s}=N_{\rm th}$, in which we see the stationary point corresponding to $\ket{-\alpha}$ state becomes unstable. The value $N_{\rm th}$ can be considered as a rough criterion of the signal strength required for the complete phase locking. Accordingly, the locking signal strength required for phase locking is higher for JPOs with higher dimensionless nonlinearity $\gamma^\prime$ and higher oscillation amplitude $\alpha$.

\section{Measured power dependence of phase locking}
\begin{figure}[t]
  \begin{center} 
    \includegraphics[width=\linewidth]{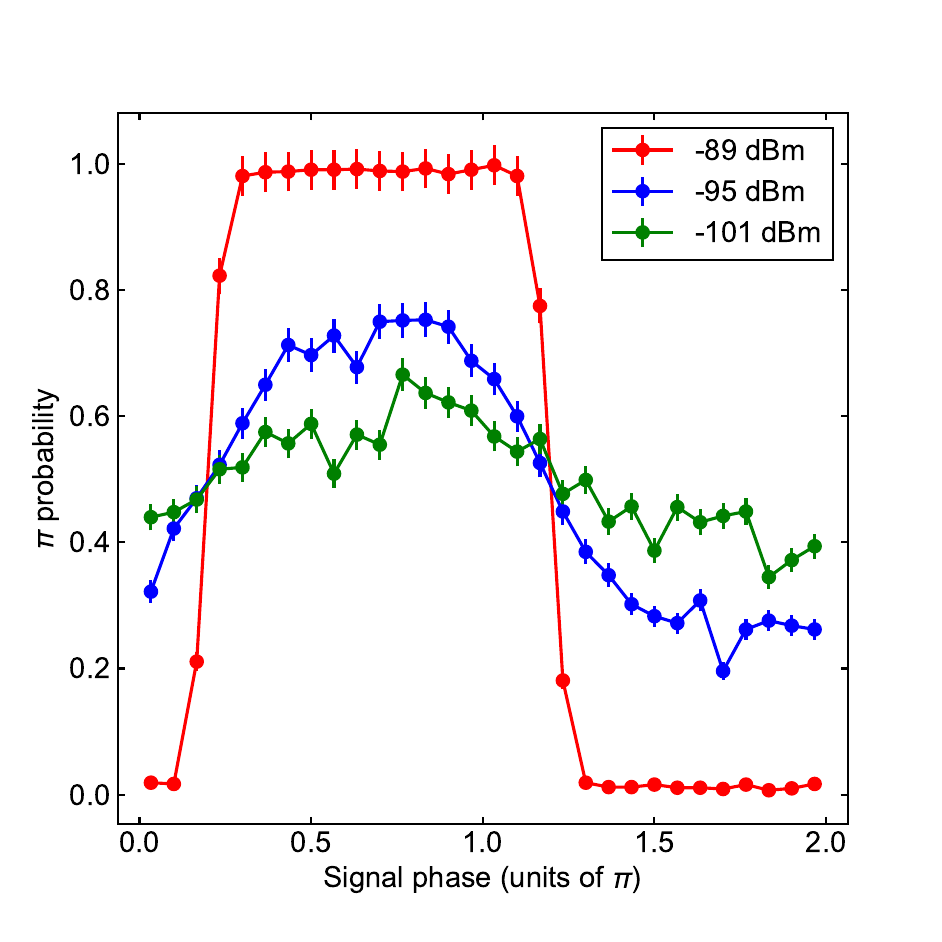}
    \caption{Phase locking measured at different signal powers. The horizontal and vertical axes show the signal phase relative to the pump and the probability of obtaining a $\pi$ state, respectively. For each signal powers, the horizontal axis is offset to compensate phase delay due to a mechanical switch for the attenuator.  The power of the locking signal is $-89$~(red), $-95$~(blue), and $-101$~(green)~dBm, respectively, and the error bars of the data points show statistical uncertainties.} 
    \label{fig:sig_pow}
  \end{center}
\end{figure}

Figure~\ref{fig:sig_pow} shows the signal-phase dependence of the probability obtaining a $\pi$ state measured at the signal power of $-89$, $-95$, and $-101$~dBm. The signal power is adjusted using a variable attenuator, and the other experimental conditions are the same as in Fig.~\ref{fig5}(c).
Compared to the measurement at $-89$~dBm, in which the output states of the JPO is fully locked by the locking signal and the locking error is $\sim$1\%, the JPO is only partially locked at lower signal powers.

\bibliography{jpo}

\providecommand{\noopsort}[1]{}\providecommand{\singleletter}[1]{#1}%
\begin{thebibliography}{41}%
\makeatletter
\providecommand \@ifxundefined [1]{%
 \@ifx{#1\undefined}
}%
\providecommand \@ifnum [1]{%
 \ifnum #1\expandafter \@firstoftwo
 \else \expandafter \@secondoftwo
 \fi
}%
\providecommand \@ifx [1]{%
 \ifx #1\expandafter \@firstoftwo
 \else \expandafter \@secondoftwo
 \fi
}%
\providecommand \natexlab [1]{#1}%
\providecommand \enquote  [1]{``#1''}%
\providecommand \bibnamefont  [1]{#1}%
\providecommand \bibfnamefont [1]{#1}%
\providecommand \citenamefont [1]{#1}%
\providecommand \href@noop [0]{\@secondoftwo}%
\providecommand \href [0]{\begingroup \@sanitize@url \@href}%
\providecommand \@href[1]{\@@startlink{#1}\@@href}%
\providecommand \@@href[1]{\endgroup#1\@@endlink}%
\providecommand \@sanitize@url [0]{\catcode `\\12\catcode `\$12\catcode
  `\&12\catcode `\#12\catcode `\^12\catcode `\_12\catcode `\%12\relax}%
\providecommand \@@startlink[1]{}%
\providecommand \@@endlink[0]{}%
\providecommand \url  [0]{\begingroup\@sanitize@url \@url }%
\providecommand \@url [1]{\endgroup\@href {#1}{\urlprefix }}%
\providecommand \urlprefix  [0]{URL }%
\providecommand \Eprint [0]{\href }%
\providecommand \doibase [0]{https://doi.org/}%
\providecommand \selectlanguage [0]{\@gobble}%
\providecommand \bibinfo  [0]{\@secondoftwo}%
\providecommand \bibfield  [0]{\@secondoftwo}%
\providecommand \translation [1]{[#1]}%
\providecommand \BibitemOpen [0]{}%
\providecommand \bibitemStop [0]{}%
\providecommand \bibitemNoStop [0]{.\EOS\space}%
\providecommand \EOS [0]{\spacefactor3000\relax}%
\providecommand \BibitemShut  [1]{\csname bibitem#1\endcsname}%
\let\auto@bib@innerbib\@empty
\bibitem [{\citenamefont {Nayfeh}\ and\ \citenamefont
  {Mook}(1995)}]{Nayfeh_book}%
  \BibitemOpen
  \bibfield  {author} {\bibinfo {author} {\bibfnamefont {A.~H.}\ \bibnamefont
  {Nayfeh}}\ and\ \bibinfo {author} {\bibfnamefont {D.~T.}\ \bibnamefont
  {Mook}},\ }\href@noop {} {\emph {\bibinfo {title} {Nonlinear Oscillations}}}\
  (\bibinfo  {publisher} {Wiley-VCH},\ \bibinfo {year} {1995})\BibitemShut
  {NoStop}%
\bibitem [{\citenamefont {Strogatz}(2000)}]{Strogatz_book}%
  \BibitemOpen
  \bibfield  {author} {\bibinfo {author} {\bibfnamefont {S.~H.}\ \bibnamefont
  {Strogatz}},\ }\href@noop {} {\emph {\bibinfo {title} {Nonlinear Dynamics and
  Chaos: With Applications To Physics, Biology, Chemistry, And Engineering
  (Studies in Nonlinearity)}}}\ (\bibinfo  {publisher} {CRC Press},\ \bibinfo
  {year} {2000})\BibitemShut {NoStop}%
\bibitem [{\citenamefont {Giordmaine}\ and\ \citenamefont
  {Miller}(1965)}]{Giordmaine1965}%
  \BibitemOpen
  \bibfield  {author} {\bibinfo {author} {\bibfnamefont {J.~A.}\ \bibnamefont
  {Giordmaine}}\ and\ \bibinfo {author} {\bibfnamefont {R.~C.}\ \bibnamefont
  {Miller}},\ }\bibfield  {title} {\bibinfo {title} {Tunable coherent
  parametric oscillation in {LiNbO$_3$} at optical frequencies},\ }\href@noop
  {} {\bibfield  {journal} {\bibinfo  {journal} {Phys. Rev. Lett.}\ }\textbf
  {\bibinfo {volume} {14}},\ \bibinfo {pages} {973} (\bibinfo {year}
  {1965})}\BibitemShut {NoStop}%
\bibitem [{\citenamefont {Turner}\ \emph {et~al.}(1998)\citenamefont {Turner},
  \citenamefont {Miller}, \citenamefont {Hartwell}, \citenamefont {MacDonald},
  \citenamefont {Strogatz},\ and\ \citenamefont {Adams}}]{Turner1998}%
  \BibitemOpen
  \bibfield  {author} {\bibinfo {author} {\bibfnamefont {K.~L.}\ \bibnamefont
  {Turner}}, \bibinfo {author} {\bibfnamefont {S.~A.}\ \bibnamefont {Miller}},
  \bibinfo {author} {\bibfnamefont {P.~G.}\ \bibnamefont {Hartwell}}, \bibinfo
  {author} {\bibfnamefont {N.~C.}\ \bibnamefont {MacDonald}}, \bibinfo {author}
  {\bibfnamefont {S.~H.}\ \bibnamefont {Strogatz}},\ and\ \bibinfo {author}
  {\bibfnamefont {S.~G.}\ \bibnamefont {Adams}},\ }\bibfield  {title} {\bibinfo
  {title} {Five parametric resonances in a microelectromechanical system},\
  }\href@noop {} {\bibfield  {journal} {\bibinfo  {journal} {Nature}\ }\textbf
  {\bibinfo {volume} {396}},\ \bibinfo {pages} {149} (\bibinfo {year}
  {1998})}\BibitemShut {NoStop}%
\bibitem [{\citenamefont {Wilson}\ \emph {et~al.}(2010)\citenamefont {Wilson},
  \citenamefont {Duty}, \citenamefont {Sandberg}, \citenamefont {Persson},
  \citenamefont {Shumeiko},\ and\ \citenamefont {Delsing}}]{Wilson2010}%
  \BibitemOpen
  \bibfield  {author} {\bibinfo {author} {\bibfnamefont {C.~M.}\ \bibnamefont
  {Wilson}}, \bibinfo {author} {\bibfnamefont {T.}~\bibnamefont {Duty}},
  \bibinfo {author} {\bibfnamefont {M.}~\bibnamefont {Sandberg}}, \bibinfo
  {author} {\bibfnamefont {F.}~\bibnamefont {Persson}}, \bibinfo {author}
  {\bibfnamefont {V.}~\bibnamefont {Shumeiko}},\ and\ \bibinfo {author}
  {\bibfnamefont {P.}~\bibnamefont {Delsing}},\ }\bibfield  {title} {\bibinfo
  {title} {Photon generation in an electromagnetic cavity with a time-dependent
  boundary},\ }\href@noop {} {\bibfield  {journal} {\bibinfo  {journal} {Phys.
  Rev. Lett.}\ }\textbf {\bibinfo {volume} {105}},\ \bibinfo {pages} {233907}
  (\bibinfo {year} {2010})}\BibitemShut {NoStop}%
\bibitem [{\citenamefont {Goto}(1959)}]{Goto1959}%
  \BibitemOpen
  \bibfield  {author} {\bibinfo {author} {\bibfnamefont {E.}~\bibnamefont
  {Goto}},\ }\bibfield  {title} {\bibinfo {title} {The parametron, a digital
  computing element which utilizes parametric oscillation},\ }\href@noop {}
  {\bibfield  {journal} {\bibinfo  {journal} {Proc. Inst. Radio Engrs.}\
  }\textbf {\bibinfo {volume} {47}},\ \bibinfo {pages} {1304} (\bibinfo {year}
  {1959})}\BibitemShut {NoStop}%
\bibitem [{\citenamefont {Boyd}(2020)}]{Boyd_book}%
  \BibitemOpen
  \bibfield  {author} {\bibinfo {author} {\bibfnamefont {R.~W.}\ \bibnamefont
  {Boyd}},\ }\href@noop {} {\emph {\bibinfo {title} {Nonlinear Optics, 3rd
  Edition}}}\ (\bibinfo  {publisher} {Academic Press},\ \bibinfo {year}
  {2020})\BibitemShut {NoStop}%
\bibitem [{\citenamefont {Castellanos-Beltran}\ \emph
  {et~al.}(2008)\citenamefont {Castellanos-Beltran}, \citenamefont {Irwin},
  \citenamefont {Hilton}, \citenamefont {Vale},\ and\ \citenamefont
  {Lehnert}}]{Beltran2008}%
  \BibitemOpen
  \bibfield  {author} {\bibinfo {author} {\bibfnamefont {M.~A.}\ \bibnamefont
  {Castellanos-Beltran}}, \bibinfo {author} {\bibfnamefont {K.~D.}\
  \bibnamefont {Irwin}}, \bibinfo {author} {\bibfnamefont {G.~C.}\ \bibnamefont
  {Hilton}}, \bibinfo {author} {\bibfnamefont {L.~R.}\ \bibnamefont {Vale}},\
  and\ \bibinfo {author} {\bibfnamefont {K.~W.}\ \bibnamefont {Lehnert}},\
  }\bibfield  {title} {\bibinfo {title} {Amplification and squeezing of quantum
  noise with a tunable {Josephson} metamaterial},\ }\href@noop {} {\bibfield
  {journal} {\bibinfo  {journal} {Nature Phys.}\ }\textbf {\bibinfo {volume}
  {4}},\ \bibinfo {pages} {929} (\bibinfo {year} {2008})}\BibitemShut {NoStop}%
\bibitem [{\citenamefont {Wustmann}\ and\ \citenamefont
  {Shumeiko}(2017)}]{Wustmann2017}%
  \BibitemOpen
  \bibfield  {author} {\bibinfo {author} {\bibfnamefont {W.}~\bibnamefont
  {Wustmann}}\ and\ \bibinfo {author} {\bibfnamefont {V.}~\bibnamefont
  {Shumeiko}},\ }\bibfield  {title} {\bibinfo {title} {Nondegenerate parametric
  resonance in a tunable superconducting cavity},\ }\href
  {https://link.aps.org/doi/10.1103/PhysRevApplied.8.024018} {\bibfield
  {journal} {\bibinfo  {journal} {Phys. Rev. Applied}\ }\textbf {\bibinfo
  {volume} {8}},\ \bibinfo {pages} {024018} (\bibinfo {year}
  {2017})}\BibitemShut {NoStop}%
\bibitem [{\citenamefont {Bengtsson}\ \emph {et~al.}(2018)\citenamefont
  {Bengtsson}, \citenamefont {Krantz}, \citenamefont {Simoen}, \citenamefont
  {Svensson}, \citenamefont {Schneider}, \citenamefont {Shumeiko},
  \citenamefont {Delsing},\ and\ \citenamefont {Bylander}}]{Bengtsson2018}%
  \BibitemOpen
  \bibfield  {author} {\bibinfo {author} {\bibfnamefont {A.}~\bibnamefont
  {Bengtsson}}, \bibinfo {author} {\bibfnamefont {P.}~\bibnamefont {Krantz}},
  \bibinfo {author} {\bibfnamefont {M.}~\bibnamefont {Simoen}}, \bibinfo
  {author} {\bibfnamefont {I.-M.}\ \bibnamefont {Svensson}}, \bibinfo {author}
  {\bibfnamefont {B.}~\bibnamefont {Schneider}}, \bibinfo {author}
  {\bibfnamefont {V.}~\bibnamefont {Shumeiko}}, \bibinfo {author}
  {\bibfnamefont {P.}~\bibnamefont {Delsing}},\ and\ \bibinfo {author}
  {\bibfnamefont {J.}~\bibnamefont {Bylander}},\ }\bibfield  {title} {\bibinfo
  {title} {Nondegenerate parametric oscillations in a tunable superconducting
  resonator},\ }\href {https://doi.org/10.1103/PhysRevB.97.144502} {\bibfield
  {journal} {\bibinfo  {journal} {Phys. Rev. B}\ }\textbf {\bibinfo {volume}
  {97}},\ \bibinfo {pages} {144502} (\bibinfo {year} {2018})}\BibitemShut
  {NoStop}%
\bibitem [{\citenamefont {Svensson}\ \emph {et~al.}(2017)\citenamefont
  {Svensson}, \citenamefont {Bengtsson}, \citenamefont {Krantz}, \citenamefont
  {Bylander}, \citenamefont {Shumeiko},\ and\ \citenamefont
  {Delsing}}]{Svensson2017}%
  \BibitemOpen
  \bibfield  {author} {\bibinfo {author} {\bibfnamefont {I.-M.}\ \bibnamefont
  {Svensson}}, \bibinfo {author} {\bibfnamefont {A.}~\bibnamefont {Bengtsson}},
  \bibinfo {author} {\bibfnamefont {P.}~\bibnamefont {Krantz}}, \bibinfo
  {author} {\bibfnamefont {J.}~\bibnamefont {Bylander}}, \bibinfo {author}
  {\bibfnamefont {V.}~\bibnamefont {Shumeiko}},\ and\ \bibinfo {author}
  {\bibfnamefont {P.}~\bibnamefont {Delsing}},\ }\bibfield  {title} {\bibinfo
  {title} {Period-tripling subharmonic oscillations in a driven superconducting
  resonator},\ }\href {https://doi.org/10.1103/PhysRevB.96.174503} {\bibfield
  {journal} {\bibinfo  {journal} {Phys. Rev. B}\ }\textbf {\bibinfo {volume}
  {96}},\ \bibinfo {pages} {174503} (\bibinfo {year} {2017})}\BibitemShut
  {NoStop}%
\bibitem [{\citenamefont {Svensson}\ \emph {et~al.}(2018)\citenamefont
  {Svensson}, \citenamefont {Bengtsson}, \citenamefont {Bylander},
  \citenamefont {Shumeiko},\ and\ \citenamefont {Delsing}}]{Svensson2018}%
  \BibitemOpen
  \bibfield  {author} {\bibinfo {author} {\bibfnamefont {I.-M.}\ \bibnamefont
  {Svensson}}, \bibinfo {author} {\bibfnamefont {A.}~\bibnamefont {Bengtsson}},
  \bibinfo {author} {\bibfnamefont {J.}~\bibnamefont {Bylander}}, \bibinfo
  {author} {\bibfnamefont {V.}~\bibnamefont {Shumeiko}},\ and\ \bibinfo
  {author} {\bibfnamefont {P.}~\bibnamefont {Delsing}},\ }\bibfield  {title}
  {\bibinfo {title} {Period multiplication in a parametrically driven
  superconducting resonator},\ }\href {https://doi.org/10.1063/1.5026974}
  {\bibfield  {journal} {\bibinfo  {journal} {Appl. Phys. Lett.}\ }\textbf
  {\bibinfo {volume} {113}},\ \bibinfo {pages} {022602} (\bibinfo {year}
  {2018})}\BibitemShut {NoStop}%
\bibitem [{\citenamefont {Goto}(2016{\natexlab{a}})}]{Goto2016_scirep}%
  \BibitemOpen
  \bibfield  {author} {\bibinfo {author} {\bibfnamefont {H.}~\bibnamefont
  {Goto}},\ }\bibfield  {title} {\bibinfo {title} {Bifurcation-based adiabatic
  quantum computation with a nonlinear oscillator network},\ }\href@noop {}
  {\bibfield  {journal} {\bibinfo  {journal} {Sci. Rep.}\ }\textbf {\bibinfo
  {volume} {6}},\ \bibinfo {pages} {21686} (\bibinfo {year}
  {2016}{\natexlab{a}})}\BibitemShut {NoStop}%
\bibitem [{\citenamefont {{G. Kirchmair, B. Vlastakis, Z. Leghtas, S. E. Nigg,
  H. Paik, E. Ginossar, M. Mirrahimi, L. Frunzio, S. M. Girvin and R. J.
  Schoelkopf}}(2013)}]{Kirchmair2013}%
  \BibitemOpen
  \bibfield  {author} {\bibinfo {author} {\bibnamefont {{G. Kirchmair, B.
  Vlastakis, Z. Leghtas, S. E. Nigg, H. Paik, E. Ginossar, M. Mirrahimi, L.
  Frunzio, S. M. Girvin and R. J. Schoelkopf}}},\ }\bibfield  {title} {\bibinfo
  {title} {Observation of quantum state collapse and revival due to the
  single-photon {Kerr} effect},\ }\href@noop {} {\bibfield  {journal} {\bibinfo
   {journal} {Nature}\ }\textbf {\bibinfo {volume} {495}},\ \bibinfo {pages}
  {205} (\bibinfo {year} {2013})}\BibitemShut {NoStop}%
\bibitem [{\citenamefont {{S. Puri, S. Boutin and A. Blais}}(2017)}]{Puri2017}%
  \BibitemOpen
  \bibfield  {author} {\bibinfo {author} {\bibnamefont {{S. Puri, S. Boutin and
  A. Blais}}},\ }\bibfield  {title} {\bibinfo {title} {Engineering the quantum
  states of light in a {Kerr}-nonlinear resonator by two-photon driving},\
  }\href@noop {} {\bibfield  {journal} {\bibinfo  {journal} {npj Quantum
  Information}\ }\textbf {\bibinfo {volume} {3}},\ \bibinfo {pages} {18}
  (\bibinfo {year} {2017})}\BibitemShut {NoStop}%
\bibitem [{\citenamefont {Goto}\ \emph {et~al.}(2019)\citenamefont {Goto},
  \citenamefont {Lin}, \citenamefont {Yamamoto},\ and\ \citenamefont
  {Nakamura}}]{Goto2019}%
  \BibitemOpen
  \bibfield  {author} {\bibinfo {author} {\bibfnamefont {H.}~\bibnamefont
  {Goto}}, \bibinfo {author} {\bibfnamefont {Z.}~\bibnamefont {Lin}}, \bibinfo
  {author} {\bibfnamefont {T.}~\bibnamefont {Yamamoto}},\ and\ \bibinfo
  {author} {\bibfnamefont {Y.}~\bibnamefont {Nakamura}},\ }\bibfield  {title}
  {\bibinfo {title} {On-demand generation of traveling cat states using a
  parametric oscillator},\ }\href@noop {} {\bibfield  {journal} {\bibinfo
  {journal} {Phys. Rev. A}\ }\textbf {\bibinfo {volume} {99}},\ \bibinfo
  {pages} {023838} (\bibinfo {year} {2019})}\BibitemShut {NoStop}%
\bibitem [{\citenamefont {Goto}(2016{\natexlab{b}})}]{Goto2016_pra}%
  \BibitemOpen
  \bibfield  {author} {\bibinfo {author} {\bibfnamefont {H.}~\bibnamefont
  {Goto}},\ }\bibfield  {title} {\bibinfo {title} {Universal quantum
  computation with a nonlinear oscillator network},\ }\href@noop {} {\bibfield
  {journal} {\bibinfo  {journal} {Phys. Rev. A}\ }\textbf {\bibinfo {volume}
  {93}},\ \bibinfo {pages} {050301} (\bibinfo {year}
  {2016}{\natexlab{b}})}\BibitemShut {NoStop}%
\bibitem [{\citenamefont {{S. E. Nigg, N. L\"{o}rch and R. P.
  Tiwari}}(2017)}]{Nigg2017}%
  \BibitemOpen
  \bibfield  {author} {\bibinfo {author} {\bibnamefont {{S. E. Nigg, N.
  L\"{o}rch and R. P. Tiwari}}},\ }\bibfield  {title} {\bibinfo {title} {Robust
  quantum optimizer with full connectivity},\ }\href@noop {} {\bibfield
  {journal} {\bibinfo  {journal} {Sci. Adv.}\ }\textbf {\bibinfo {volume}
  {3}},\ \bibinfo {pages} {e1602273} (\bibinfo {year} {2017})}\BibitemShut
  {NoStop}%
\bibitem [{\citenamefont {{S. Puri, C. K. Andersen, A. L. Grimsmo and A.
  Blais}}(2017)}]{Puri2017b}%
  \BibitemOpen
  \bibfield  {author} {\bibinfo {author} {\bibnamefont {{S. Puri, C. K.
  Andersen, A. L. Grimsmo and A. Blais}}},\ }\bibfield  {title} {\bibinfo
  {title} {Quantum annealing with all-to-all connected nonlinear oscillators},\
  }\href@noop {} {\bibfield  {journal} {\bibinfo  {journal} {Nat. Commun.}\
  }\textbf {\bibinfo {volume} {8}},\ \bibinfo {pages} {15785} (\bibinfo {year}
  {2017})}\BibitemShut {NoStop}%
\bibitem [{\citenamefont {Zhao}\ \emph {et~al.}(2018)\citenamefont {Zhao},
  \citenamefont {Jin}, \citenamefont {Xu}, \citenamefont {Tan}, \citenamefont
  {Yu},\ and\ \citenamefont {Yu}}]{Zhao2018}%
  \BibitemOpen
  \bibfield  {author} {\bibinfo {author} {\bibfnamefont {P.}~\bibnamefont
  {Zhao}}, \bibinfo {author} {\bibfnamefont {Z.}~\bibnamefont {Jin}}, \bibinfo
  {author} {\bibfnamefont {P.}~\bibnamefont {Xu}}, \bibinfo {author}
  {\bibfnamefont {X.}~\bibnamefont {Tan}}, \bibinfo {author} {\bibfnamefont
  {H.}~\bibnamefont {Yu}},\ and\ \bibinfo {author} {\bibfnamefont
  {Y.}~\bibnamefont {Yu}},\ }\bibfield  {title} {\bibinfo {title} {Two-photon
  driven {Kerr} resonator for quantum annealing with three-dimensional circuit
  {QED}},\ }\href@noop {} {\bibfield  {journal} {\bibinfo  {journal} {Phys.
  Rev. Applied}\ }\textbf {\bibinfo {volume} {10}},\ \bibinfo {pages} {024019}
  (\bibinfo {year} {2018})}\BibitemShut {NoStop}%
\bibitem [{\citenamefont {Goto}(2019)}]{Goto2019_jpsj}%
  \BibitemOpen
  \bibfield  {author} {\bibinfo {author} {\bibfnamefont {H.}~\bibnamefont
  {Goto}},\ }\bibfield  {title} {\bibinfo {title} {Quantum computation based on
  quantum adiabatic bifurcations of {Kerr}-nonlinear parametric oscillators},\
  }\href@noop {} {\bibfield  {journal} {\bibinfo  {journal} {J. Phys. Soc.
  Jpn.}\ }\textbf {\bibinfo {volume} {88}},\ \bibinfo {pages} {061015}
  (\bibinfo {year} {2019})}\BibitemShut {NoStop}%
\bibitem [{\citenamefont {Goto}\ and\ \citenamefont {Kanao}(2020)}]{Goto2020}%
  \BibitemOpen
  \bibfield  {author} {\bibinfo {author} {\bibfnamefont {H.}~\bibnamefont
  {Goto}}\ and\ \bibinfo {author} {\bibfnamefont {T.}~\bibnamefont {Kanao}},\
  }\href@noop {} {\bibinfo {title} {Excited-state adiabatic quantum computation
  started with vacuum states}} (\bibinfo {year} {2020}),\ \Eprint
  {https://arxiv.org/abs/arXiv:2005.07511} {arXiv:2005.07511} \BibitemShut
  {NoStop}%
\bibitem [{\citenamefont {Kanao}\ and\ \citenamefont {Goto}(2020)}]{Kanao2020}%
  \BibitemOpen
  \bibfield  {author} {\bibinfo {author} {\bibfnamefont {T.}~\bibnamefont
  {Kanao}}\ and\ \bibinfo {author} {\bibfnamefont {H.}~\bibnamefont {Goto}},\
  }\href@noop {} {\bibinfo {title} {High-accuracy {Ising} machine using
  {Kerr}-nonlinear parametric oscillators with local four-body interactions}}
  (\bibinfo {year} {2020}),\ \Eprint {https://arxiv.org/abs/arXiv:2005.13819}
  {arXiv:2005.13819} \BibitemShut {NoStop}%
\bibitem [{\citenamefont {Ding}\ \emph {et~al.}(2017)\citenamefont {Ding},
  \citenamefont {Maslennikov}, \citenamefont {Habl\"utzel}, \citenamefont
  {Loh},\ and\ \citenamefont {Matsukevich}}]{Ding2017}%
  \BibitemOpen
  \bibfield  {author} {\bibinfo {author} {\bibfnamefont {S.}~\bibnamefont
  {Ding}}, \bibinfo {author} {\bibfnamefont {G.}~\bibnamefont {Maslennikov}},
  \bibinfo {author} {\bibfnamefont {R.}~\bibnamefont {Habl\"utzel}}, \bibinfo
  {author} {\bibfnamefont {H.}~\bibnamefont {Loh}},\ and\ \bibinfo {author}
  {\bibfnamefont {D.}~\bibnamefont {Matsukevich}},\ }\bibfield  {title}
  {\bibinfo {title} {Quantum parametric oscillator with trapped ions},\
  }\href@noop {} {\bibfield  {journal} {\bibinfo  {journal} {Phys. Rev. Lett.}\
  }\textbf {\bibinfo {volume} {119}},\ \bibinfo {pages} {150404} (\bibinfo
  {year} {2017})}\BibitemShut {NoStop}%
\bibitem [{\citenamefont {Wang}\ \emph {et~al.}(2019)\citenamefont {Wang},
  \citenamefont {Pechal}, \citenamefont {Wollack}, \citenamefont
  {Arrangoiz-Arriola}, \citenamefont {Gao}, \citenamefont {Lee},\ and\
  \citenamefont {Safavi-Naeini}}]{Wang2019}%
  \BibitemOpen
  \bibfield  {author} {\bibinfo {author} {\bibfnamefont {Z.}~\bibnamefont
  {Wang}}, \bibinfo {author} {\bibfnamefont {M.}~\bibnamefont {Pechal}},
  \bibinfo {author} {\bibfnamefont {E.~A.}\ \bibnamefont {Wollack}}, \bibinfo
  {author} {\bibfnamefont {P.}~\bibnamefont {Arrangoiz-Arriola}}, \bibinfo
  {author} {\bibfnamefont {M.}~\bibnamefont {Gao}}, \bibinfo {author}
  {\bibfnamefont {N.~R.}\ \bibnamefont {Lee}},\ and\ \bibinfo {author}
  {\bibfnamefont {A.~H.}\ \bibnamefont {Safavi-Naeini}},\ }\bibfield  {title}
  {\bibinfo {title} {Quantum dynamics of a few-photon parametric oscillator},\
  }\href@noop {} {\bibfield  {journal} {\bibinfo  {journal} {Phys. Rev. X}\
  }\textbf {\bibinfo {volume} {9}},\ \bibinfo {pages} {021049} (\bibinfo {year}
  {2019})}\BibitemShut {NoStop}%
\bibitem [{\citenamefont {Grimm}\ \emph {et~al.}(2020)\citenamefont {Grimm},
  \citenamefont {Frattini}, \citenamefont {Puri}, \citenamefont {Mundhada},
  \citenamefont {Touzard}, \citenamefont {Mirrahimi}, \citenamefont {Girvin},
  \citenamefont {Shankar},\ and\ \citenamefont {Devoret}}]{Grimm2020}%
  \BibitemOpen
  \bibfield  {author} {\bibinfo {author} {\bibfnamefont {A.}~\bibnamefont
  {Grimm}}, \bibinfo {author} {\bibfnamefont {N.~E.}\ \bibnamefont {Frattini}},
  \bibinfo {author} {\bibfnamefont {S.}~\bibnamefont {Puri}}, \bibinfo {author}
  {\bibfnamefont {S.~O.}\ \bibnamefont {Mundhada}}, \bibinfo {author}
  {\bibfnamefont {S.}~\bibnamefont {Touzard}}, \bibinfo {author} {\bibfnamefont
  {M.}~\bibnamefont {Mirrahimi}}, \bibinfo {author} {\bibfnamefont {S.~M.}\
  \bibnamefont {Girvin}}, \bibinfo {author} {\bibfnamefont {S.}~\bibnamefont
  {Shankar}},\ and\ \bibinfo {author} {\bibfnamefont {M.~H.}\ \bibnamefont
  {Devoret}},\ }\bibfield  {title} {\bibinfo {title} {Stabilization and
  operation of a {Kerr}-cat qubit},\ }\href
  {https://doi.org/10.1038/s41586-020-2587-z} {\bibfield  {journal} {\bibinfo
  {journal} {Nature}\ }\textbf {\bibinfo {volume} {584}},\ \bibinfo {pages}
  {205} (\bibinfo {year} {2020})}\BibitemShut {NoStop}%
\bibitem [{\citenamefont {Leib}\ \emph {et~al.}(2012)\citenamefont {Leib},
  \citenamefont {Deppe}, \citenamefont {Marx}, \citenamefont {Gross},\ and\
  \citenamefont {Hartmann}}]{Leib2012}%
  \BibitemOpen
  \bibfield  {author} {\bibinfo {author} {\bibfnamefont {M.}~\bibnamefont
  {Leib}}, \bibinfo {author} {\bibfnamefont {F.}~\bibnamefont {Deppe}},
  \bibinfo {author} {\bibfnamefont {A.}~\bibnamefont {Marx}}, \bibinfo {author}
  {\bibfnamefont {R.}~\bibnamefont {Gross}},\ and\ \bibinfo {author}
  {\bibfnamefont {M.~J.}\ \bibnamefont {Hartmann}},\ }\bibfield  {title}
  {\bibinfo {title} {Networks of nonlinear superconducting transmission line
  resonators},\ }\href@noop {} {\bibfield  {journal} {\bibinfo  {journal} {New
  Journal of Physics}\ }\textbf {\bibinfo {volume} {14}},\ \bibinfo {pages}
  {075024} (\bibinfo {year} {2012})}\BibitemShut {NoStop}%
\bibitem [{\citenamefont {Muppalla}\ \emph {et~al.}(2018)\citenamefont
  {Muppalla}, \citenamefont {Gargiulo}, \citenamefont {Mirzaei}, \citenamefont
  {Venkatesh}, \citenamefont {Juan}, \citenamefont {Gr\"unhaupt}, \citenamefont
  {Pop},\ and\ \citenamefont {Kirchmair}}]{Muppalla2018}%
  \BibitemOpen
  \bibfield  {author} {\bibinfo {author} {\bibfnamefont {P.~R.}\ \bibnamefont
  {Muppalla}}, \bibinfo {author} {\bibfnamefont {O.}~\bibnamefont {Gargiulo}},
  \bibinfo {author} {\bibfnamefont {S.~I.}\ \bibnamefont {Mirzaei}}, \bibinfo
  {author} {\bibfnamefont {B.~P.}\ \bibnamefont {Venkatesh}}, \bibinfo {author}
  {\bibfnamefont {M.~L.}\ \bibnamefont {Juan}}, \bibinfo {author}
  {\bibfnamefont {L.}~\bibnamefont {Gr\"unhaupt}}, \bibinfo {author}
  {\bibfnamefont {I.~M.}\ \bibnamefont {Pop}},\ and\ \bibinfo {author}
  {\bibfnamefont {G.}~\bibnamefont {Kirchmair}},\ }\bibfield  {title} {\bibinfo
  {title} {Bistability in a mesoscopic {Josephson} junction array resonator},\
  }\href@noop {} {\bibfield  {journal} {\bibinfo  {journal} {Phys. Rev. B}\
  }\textbf {\bibinfo {volume} {97}},\ \bibinfo {pages} {024518} (\bibinfo
  {year} {2018})}\BibitemShut {NoStop}%
\bibitem [{\citenamefont {Andersen}\ \emph {et~al.}(2020)\citenamefont
  {Andersen}, \citenamefont {Kamal}, \citenamefont {Masluk}, \citenamefont
  {Pop}, \citenamefont {Blais},\ and\ \citenamefont {Devoret}}]{Andersen2020}%
  \BibitemOpen
  \bibfield  {author} {\bibinfo {author} {\bibfnamefont {C.~K.}\ \bibnamefont
  {Andersen}}, \bibinfo {author} {\bibfnamefont {A.}~\bibnamefont {Kamal}},
  \bibinfo {author} {\bibfnamefont {N.~A.}\ \bibnamefont {Masluk}}, \bibinfo
  {author} {\bibfnamefont {I.~M.}\ \bibnamefont {Pop}}, \bibinfo {author}
  {\bibfnamefont {A.}~\bibnamefont {Blais}},\ and\ \bibinfo {author}
  {\bibfnamefont {M.~H.}\ \bibnamefont {Devoret}},\ }\bibfield  {title}
  {\bibinfo {title} {Quantum versus classical switching dynamics of driven
  dissipative {Kerr} resonators},\ }\href@noop {} {\bibfield  {journal}
  {\bibinfo  {journal} {Phys. Rev. Applied}\ }\textbf {\bibinfo {volume}
  {13}},\ \bibinfo {pages} {044017} (\bibinfo {year} {2020})}\BibitemShut
  {NoStop}%
\bibitem [{\citenamefont {Yamamoto}\ \emph {et~al.}(2016)\citenamefont
  {Yamamoto}, \citenamefont {Koshino},\ and\ \citenamefont
  {Nakamura}}]{Yamamoto_book}%
  \BibitemOpen
  \bibfield  {author} {\bibinfo {author} {\bibfnamefont {T.}~\bibnamefont
  {Yamamoto}}, \bibinfo {author} {\bibfnamefont {K.}~\bibnamefont {Koshino}},\
  and\ \bibinfo {author} {\bibfnamefont {Y.}~\bibnamefont {Nakamura}},\
  }\href@noop {} {\emph {\bibinfo {title} {\emph{in} Principles and Methods of
  Quantum Information Technologies}}},\ edited by\ \bibinfo {editor}
  {\bibfnamefont {Y.}~\bibnamefont {Yamamoto}}\ and\ \bibinfo {editor}
  {\bibfnamefont {K.}~\bibnamefont {Semba}}\ (\bibinfo  {publisher} {Springer
  Japan},\ \bibinfo {year} {2016})\ Chap.~\bibinfo {chapter} {23}, pp.\
  \bibinfo {pages} {495--513}\BibitemShut {NoStop}%
\bibitem [{\citenamefont {Bourassa}\ \emph {et~al.}(2012)\citenamefont
  {Bourassa}, \citenamefont {Beaudoin}, \citenamefont {Gambetta},\ and\
  \citenamefont {Blais}}]{Bourassa12}%
  \BibitemOpen
  \bibfield  {author} {\bibinfo {author} {\bibfnamefont {J.}~\bibnamefont
  {Bourassa}}, \bibinfo {author} {\bibfnamefont {F.}~\bibnamefont {Beaudoin}},
  \bibinfo {author} {\bibfnamefont {J.~M.}\ \bibnamefont {Gambetta}},\ and\
  \bibinfo {author} {\bibfnamefont {A.}~\bibnamefont {Blais}},\ }\bibfield
  {title} {\bibinfo {title} {Josephson-junction-embedded transmission-line
  resonators: From {Kerr} medium to in-line transmon},\ }\href@noop {}
  {\bibfield  {journal} {\bibinfo  {journal} {Phys. Rev. A}\ }\textbf {\bibinfo
  {volume} {86}},\ \bibinfo {pages} {013814} (\bibinfo {year}
  {2012})}\BibitemShut {NoStop}%
\bibitem [{\citenamefont {Yurke}\ and\ \citenamefont {Buks}(2006)}]{Yurke06}%
  \BibitemOpen
  \bibfield  {author} {\bibinfo {author} {\bibfnamefont {B.}~\bibnamefont
  {Yurke}}\ and\ \bibinfo {author} {\bibfnamefont {E.}~\bibnamefont {Buks}},\
  }\bibfield  {title} {\bibinfo {title} {Performance of cavity-parametric
  amplifiers, employing {Kerr} nonlinearites, in the presence of two-photon
  loss},\ }\href@noop {} {\bibfield  {journal} {\bibinfo  {journal} {J.
  Lightwave Technol.}\ }\textbf {\bibinfo {volume} {24}},\ \bibinfo {pages}
  {5054} (\bibinfo {year} {2006})}\BibitemShut {NoStop}%
\bibitem [{\citenamefont {Frattini}\ \emph {et~al.}(2018)\citenamefont
  {Frattini}, \citenamefont {Sivak}, \citenamefont {Lingenfelter},
  \citenamefont {Shankar},\ and\ \citenamefont {Devoret}}]{Frattini2018}%
  \BibitemOpen
  \bibfield  {author} {\bibinfo {author} {\bibfnamefont {N.~E.}\ \bibnamefont
  {Frattini}}, \bibinfo {author} {\bibfnamefont {V.~V.}\ \bibnamefont {Sivak}},
  \bibinfo {author} {\bibfnamefont {A.}~\bibnamefont {Lingenfelter}}, \bibinfo
  {author} {\bibfnamefont {S.}~\bibnamefont {Shankar}},\ and\ \bibinfo {author}
  {\bibfnamefont {M.~H.}\ \bibnamefont {Devoret}},\ }\bibfield  {title}
  {\bibinfo {title} {Optimizing the nonlinearity and dissipation of a snail
  parametric amplifier for dynamic range},\ }\href
  {https://link.aps.org/doi/10.1103/PhysRevApplied.10.054020} {\bibfield
  {journal} {\bibinfo  {journal} {Phys. Rev. Appl.}\ }\textbf {\bibinfo
  {volume} {10}},\ \bibinfo {pages} {054020} (\bibinfo {year}
  {2018})}\BibitemShut {NoStop}%
\bibitem [{\citenamefont {Sivak}\ \emph {et~al.}(2019)\citenamefont {Sivak},
  \citenamefont {Frattini}, \citenamefont {Joshi}, \citenamefont
  {Lingenfelter}, \citenamefont {Shankar},\ and\ \citenamefont
  {Devoret}}]{Sivak2019}%
  \BibitemOpen
  \bibfield  {author} {\bibinfo {author} {\bibfnamefont {V.~V.}\ \bibnamefont
  {Sivak}}, \bibinfo {author} {\bibfnamefont {N.~E.}\ \bibnamefont {Frattini}},
  \bibinfo {author} {\bibfnamefont {V.~R.}\ \bibnamefont {Joshi}}, \bibinfo
  {author} {\bibfnamefont {A.}~\bibnamefont {Lingenfelter}}, \bibinfo {author}
  {\bibfnamefont {S.}~\bibnamefont {Shankar}},\ and\ \bibinfo {author}
  {\bibfnamefont {M.~H.}\ \bibnamefont {Devoret}},\ }\bibfield  {title}
  {\bibinfo {title} {Kerr-free three-wave mixing in superconducting quantum
  circuits},\ }\href
  {https://link.aps.org/doi/10.1103/PhysRevApplied.11.054060} {\bibfield
  {journal} {\bibinfo  {journal} {Phys. Rev. Appl.}\ }\textbf {\bibinfo
  {volume} {11}},\ \bibinfo {pages} {054060} (\bibinfo {year}
  {2019})}\BibitemShut {NoStop}%
\bibitem [{\citenamefont {Koshino}\ \emph {et~al.}(2013)\citenamefont
  {Koshino}, \citenamefont {Terai}, \citenamefont {Inomata}, \citenamefont
  {Yamamoto}, \citenamefont {Qiu}, \citenamefont {Wang},\ and\ \citenamefont
  {Nakamura}}]{Koshino2013}%
  \BibitemOpen
  \bibfield  {author} {\bibinfo {author} {\bibfnamefont {K.}~\bibnamefont
  {Koshino}}, \bibinfo {author} {\bibfnamefont {H.}~\bibnamefont {Terai}},
  \bibinfo {author} {\bibfnamefont {K.}~\bibnamefont {Inomata}}, \bibinfo
  {author} {\bibfnamefont {T.}~\bibnamefont {Yamamoto}}, \bibinfo {author}
  {\bibfnamefont {W.}~\bibnamefont {Qiu}}, \bibinfo {author} {\bibfnamefont
  {Z.}~\bibnamefont {Wang}},\ and\ \bibinfo {author} {\bibfnamefont
  {Y.}~\bibnamefont {Nakamura}},\ }\bibfield  {title} {\bibinfo {title}
  {Observation of the three-state dressed states in circuit quantum
  electrodynamics},\ }\href {https://doi.org/10.1103/PhysRevLett.110.263601}
  {\bibfield  {journal} {\bibinfo  {journal} {Phys. Rev. Lett.}\ }\textbf
  {\bibinfo {volume} {110}},\ \bibinfo {pages} {263601} (\bibinfo {year}
  {2013})}\BibitemShut {NoStop}%
\bibitem [{\citenamefont {Schuster}\ \emph {et~al.}(2005)\citenamefont
  {Schuster}, \citenamefont {Wallraff}, \citenamefont {Blais}, \citenamefont
  {Frunzio}, \citenamefont {Huang}, \citenamefont {Majer}, \citenamefont
  {Girvin},\ and\ \citenamefont {Schoelkopf}}]{Schuster2005}%
  \BibitemOpen
  \bibfield  {author} {\bibinfo {author} {\bibfnamefont {D.~I.}\ \bibnamefont
  {Schuster}}, \bibinfo {author} {\bibfnamefont {A.}~\bibnamefont {Wallraff}},
  \bibinfo {author} {\bibfnamefont {A.}~\bibnamefont {Blais}}, \bibinfo
  {author} {\bibfnamefont {L.}~\bibnamefont {Frunzio}}, \bibinfo {author}
  {\bibfnamefont {R.-S.}\ \bibnamefont {Huang}}, \bibinfo {author}
  {\bibfnamefont {J.}~\bibnamefont {Majer}}, \bibinfo {author} {\bibfnamefont
  {S.~M.}\ \bibnamefont {Girvin}},\ and\ \bibinfo {author} {\bibfnamefont
  {R.~J.}\ \bibnamefont {Schoelkopf}},\ }\bibfield  {title} {\bibinfo {title}
  {ac {Stark} shift and dephasing of a superconducting qubit strongly coupled
  to a cavity field},\ }\href@noop {} {\bibfield  {journal} {\bibinfo
  {journal} {Phys. Rev. Lett.}\ }\textbf {\bibinfo {volume} {94}},\ \bibinfo
  {pages} {123602} (\bibinfo {year} {2005})}\BibitemShut {NoStop}%
\bibitem [{\citenamefont {{D. I. Schuster, A. A. Houck, J. A. Schreier, A.
  Wallraff, J. M. Gambetta, A. Blais, L. Frunzio, J. Majer, B. Johnson, M. H.
  Devoret, S. M. Girvin and R. J. Schoelkopf}}(2007)}]{Schuster2007}%
  \BibitemOpen
  \bibfield  {author} {\bibinfo {author} {\bibnamefont {{D. I. Schuster, A. A.
  Houck, J. A. Schreier, A. Wallraff, J. M. Gambetta, A. Blais, L. Frunzio, J.
  Majer, B. Johnson, M. H. Devoret, S. M. Girvin and R. J. Schoelkopf}}},\
  }\bibfield  {title} {\bibinfo {title} {Resolving photon number states in a
  superconducting circuit},\ }\href@noop {} {\bibfield  {journal} {\bibinfo
  {journal} {Nature}\ }\textbf {\bibinfo {volume} {445}},\ \bibinfo {pages}
  {515} (\bibinfo {year} {2007})}\BibitemShut {NoStop}%
\bibitem [{\citenamefont {Krantz}\ \emph {et~al.}(2013)\citenamefont {Krantz},
  \citenamefont {Reshitnyk}, \citenamefont {Wustmann}, \citenamefont
  {Bylander}, \citenamefont {Gustavsson}, \citenamefont {Oliver}, \citenamefont
  {Duty}, \citenamefont {Shumeiko},\ and\ \citenamefont
  {Delsing}}]{Krantz2013}%
  \BibitemOpen
  \bibfield  {author} {\bibinfo {author} {\bibfnamefont {P.}~\bibnamefont
  {Krantz}}, \bibinfo {author} {\bibfnamefont {Y.}~\bibnamefont {Reshitnyk}},
  \bibinfo {author} {\bibfnamefont {W.}~\bibnamefont {Wustmann}}, \bibinfo
  {author} {\bibfnamefont {J.}~\bibnamefont {Bylander}}, \bibinfo {author}
  {\bibfnamefont {S.}~\bibnamefont {Gustavsson}}, \bibinfo {author}
  {\bibfnamefont {W.~D.}\ \bibnamefont {Oliver}}, \bibinfo {author}
  {\bibfnamefont {T.}~\bibnamefont {Duty}}, \bibinfo {author} {\bibfnamefont
  {V.}~\bibnamefont {Shumeiko}},\ and\ \bibinfo {author} {\bibfnamefont
  {P.}~\bibnamefont {Delsing}},\ }\bibfield  {title} {\bibinfo {title}
  {Investigation of nonlinear effects in {Josephson} parametric oscillators
  used in circuit quantum electrodynamics},\ }\href@noop {} {\bibfield
  {journal} {\bibinfo  {journal} {New Journal of Physics}\ }\textbf {\bibinfo
  {volume} {15}},\ \bibinfo {pages} {105002} (\bibinfo {year}
  {2013})}\BibitemShut {NoStop}%
\bibitem [{\citenamefont {Wustmann}\ and\ \citenamefont
  {Shumeiko}(2013)}]{Wustmann2013}%
  \BibitemOpen
  \bibfield  {author} {\bibinfo {author} {\bibfnamefont {W.}~\bibnamefont
  {Wustmann}}\ and\ \bibinfo {author} {\bibfnamefont {V.}~\bibnamefont
  {Shumeiko}},\ }\bibfield  {title} {\bibinfo {title} {Parametric resonance in
  tunable superconducting cavities},\ }\href
  {https://doi.org/10.1103/PhysRevB.87.184501} {\bibfield  {journal} {\bibinfo
  {journal} {Phys. Rev. B}\ }\textbf {\bibinfo {volume} {87}},\ \bibinfo
  {pages} {184501} (\bibinfo {year} {2013})}\BibitemShut {NoStop}%
\bibitem [{\citenamefont {Lin}\ \emph {et~al.}(2014)\citenamefont {Lin},
  \citenamefont {Inomata}, \citenamefont {Koshino}, \citenamefont {Oliver},
  \citenamefont {Nakamura}, \citenamefont {Tsai},\ and\ \citenamefont
  {Yamamoto}}]{Lin2014}%
  \BibitemOpen
  \bibfield  {author} {\bibinfo {author} {\bibfnamefont {Z.~R.}\ \bibnamefont
  {Lin}}, \bibinfo {author} {\bibfnamefont {K.}~\bibnamefont {Inomata}},
  \bibinfo {author} {\bibfnamefont {K.}~\bibnamefont {Koshino}}, \bibinfo
  {author} {\bibfnamefont {W.~D.}\ \bibnamefont {Oliver}}, \bibinfo {author}
  {\bibfnamefont {Y.}~\bibnamefont {Nakamura}}, \bibinfo {author}
  {\bibfnamefont {J.~S.}\ \bibnamefont {Tsai}},\ and\ \bibinfo {author}
  {\bibfnamefont {T.}~\bibnamefont {Yamamoto}},\ }\bibfield  {title} {\bibinfo
  {title} {Josephson parametric phase-locked oscillator and its application to
  dispersive readout of superconducting qubits},\ }\href@noop {} {\bibfield
  {journal} {\bibinfo  {journal} {Nature Commun.}\ }\textbf {\bibinfo {volume}
  {5}},\ \bibinfo {pages} {4480} (\bibinfo {year} {2014})}\BibitemShut
  {NoStop}%
\bibitem [{\citenamefont {Marthaler}\ and\ \citenamefont
  {Dykman}(2006)}]{Marthaler2006}%
  \BibitemOpen
  \bibfield  {author} {\bibinfo {author} {\bibfnamefont {M.}~\bibnamefont
  {Marthaler}}\ and\ \bibinfo {author} {\bibfnamefont {M.~I.}\ \bibnamefont
  {Dykman}},\ }\bibfield  {title} {\bibinfo {title} {Switching via quantum
  activation: A parametrically modulated oscillator},\ }\href
  {https://doi.org/10.1103/PhysRevA.73.042108} {\bibfield  {journal} {\bibinfo
  {journal} {Phys. Rev. A}\ }\textbf {\bibinfo {volume} {73}},\ \bibinfo
  {pages} {042108} (\bibinfo {year} {2006})}\BibitemShut {NoStop}%
\end{thebibliography}%

\end{document}